\documentstyle[multicol,epsf,aps,prb]{revtex}

\renewcommand{\narrowtext}{\begin{multicols}{2} \global\columnwidth20.5pc}
\renewcommand{\widetext}{\end{multicols} \global\columnwidth42.5pc}

\input{epsf.tex}

\def\top#1{\vskip #1\begin{picture}(290,80)(80,500)\thinlines \put(
65,500){\line( 1, 0){255}}\put(320,500){\line( 0, 1){
5}}\end{picture}}
\def\bottom#1{\vskip #1\begin{picture}(290,80)(80,500)\thinlines \put(
330,500){\line( 1, 0){255}}\put(330,500){\line( 0, -1){
5}}\end{picture}}

\newcommand{\ket}[1]{\left | \, #1 \right \rangle}
\newcommand{\bra}[1]{\left \langle #1 \, \right |}

\newcommand{\beq}{\begin{equation}}
\newcommand{\eeq}{\end{equation}}

\begin{document}

\draft

\preprint{IUCM97-033}

\title{Observability of counterpropagating modes at
 fractional-quantum-Hall edges}

\author{U.~Z\"ulicke and A.~H.~MacDonald}
\address{Department of Physics, Indiana University, Bloomington,
Indiana 47405}

\author{M.~D.~Johnson}
\address{Department of Physics, University of Central Florida,
Orlando, Florida 32816}

\date{\today}

\maketitle

\begin{abstract}

When the bulk filling factor is $\nu = 1 - 1/m$ with $m$ odd, at
least one counterpropagating chiral collective mode occurs
simultaneously with magnetoplasmons at the edge of
fractional-quantum-Hall samples. Initial experimental searches
for an additional mode were unsuccessful. In this paper, we
address conditions under which its observation should be
expected in experiments where the electronic system is excited
and probed by capacitive coupling. We derive realistic
expressions for the velocity of the slow counterpropagating
mode, starting from a microscopic calculation which is
simplified by a Landau-Silin-like separation between long-range
Hartree and residual interactions. The microscopic calculation
determines the stiffness of the edge to long-wavelength neutral
excitations, which fixes the slow-mode velocity, and the
effective width of the edge region, which influences the
magnetoplasmon dispersion.
 
\end{abstract}


\narrowtext

\section{Introduction}

A two-dimensional (2D) electron system in a strong transverse
magnetic field can exhibit the quantum Hall (QH)
effect.\cite{qhe-sg,qhe-tc} This effect occurs when the electron
fluid becomes incompressible\cite{ahmintro} at
magnetic-field-dependent densities. The physical origin of the
incompressibility, i.e., of an energy gap for the excitation of 
unbound particle-hole pairs, is quite different for the
integer and fractional QH effects. In the integer case, the
incompressibility arises from Landau quantization of the
kinetic energy of a charged 2D particle in a transverse magnetic
field, while in the fractional case it is a consequence of
electron-electron interactions. In both cases, however, the
only low-lying excitations are localized at the boundary of
the QH sample. In a magnetic field, collective modes, known as
edge magnetoplasmons\cite{emp} (EMP), occur at the edge of a 2D
electron system even when the bulk is compressible. Outside of
the QH regime, however, these modes have a finite
lifetime\cite{emp} for decay into incoherent particle-hole
excitations and are most aptly described using a hydrodynamic
picture. In the QH regime, provided that the edge of the 2D
electron system is sufficiently sharp,\cite{smooth} the
microscopic physics simplifies and there is no particle-hole
continuum into which the modes can decay. Generalizations of
models familiar from the study of one-dimensional (1D) electron
systems\cite{voit:reprog:94} can then be used to provide a fully
microscopic description of integer\cite{bih:prb:82} and
fractional\cite{wen:int:92,wen:adv:95} QH edges. In these
models, EMP appear as free Bose particles in the bosonized
description of a chiral 1D electron gas.

\begin{figure}[hbt]
\epsfxsize3.5in
\centerline{\epsffile{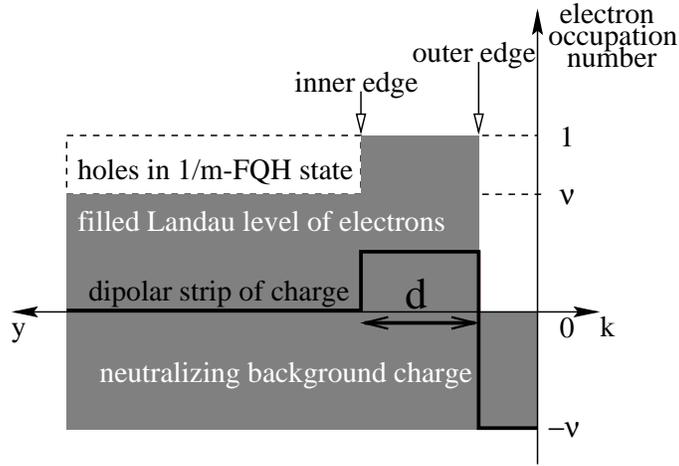}}
\vspace{0.2cm}
\caption{Schematic electron-occupation-number profile at the edge
of a fractional-QH sample for filling factor $\nu=1-1/m$. This
picture (see Ref.~\protect\onlinecite{ahm:prl:90}) is based on
the use of particle-hole conjugation
(Refs.~\protect\onlinecite{smg:prb:84,ahm:dbm:prb:85})
to understand the bulk $\nu = 1 - 1/m$ fractional QH effect.
The states are most conveniently described in terms of holes in
a filled Landau level. Two fractional-QH strips are formed by
the {\em holes\/}: an inner strip with filling factor
$\nu^{\text{(i)}} = 1-\nu \equiv 1/m$ and an outer strip with
filling factor $\nu^{\text{(o)}} = 1$.  The two strips are
separated by a distance $d$. The abscissa is the wave vector $k$
parallel to the edge. In the Landau gauge, a state having wave
vector $k$ is localized at a position $y$ perpendicular to the
edge which is proportional to $k$. Here, we measure $y$ from the
physical boundary of the sample inwards. If the 2D electron
system is placed in a coplanar neutralizing background charge
with a sharp edge, the total charge density will be negative
between inner and outer hole strips and positive inside the
outer hole strip. The schematic illustration of the resulting
dipolar strip of charge is unrealistic in its depiction of the
variation of charge density across the edge. The density profiles
at the edges of both hole strips vary on a magnetic-length scale.
Accounting for this in our calculations requires only the
introduction of appropriate form factors.}
\label{phconj}
\end{figure}

In this work, we consider the edge of a 2D electron system in
the regime where the fractional QH effect occurs, i.e., for
filling factors $\nu < 1$ and $\nu$ equal to one of the filling
factors at which the bulk of the 2D system is incompressible.
The fractional quantum Hall effect is most easily understood for 
$\nu = 1/m$ where $m=1, 3, 5, \dots$. For these values of $\nu$
and for a confining potential that is sharp enough to prevent
edge reconstruction,\cite{ahm:aust:93,wen:prb:94} a single
branch of bosonized excitations
occurs.\cite{ahm:prl:90,wen:int:92} These are EMP modes, which,
in this case, have an especially simple microscopic description.
Since the magnetic field breaks time-reversal symmetry, EMP
modes propagate along the edge in one direction only; they are
chiral. In general, however, the edge can support more than one
branch of chiral edge excitations, and some of these can
propagate in the opposite direction. For example,
counterpropagating modes can occur even at integer filling
factors when an edge reconstruction takes place.\cite{wen:prb:94}
Here we study in detail the case of bulk filling factors
$\nu = 1 - 1/m$ for which both\cite{ahm:prl:90,wen:int:92}
microscopic theory and phenomenological considerations suggest
that, even when the edge is sharp, at least one
counterpropagating mode exists in addition to the EMP mode.
For short-range electron-electron interactions, the two
collective modes consist of an outer mode similar to the $\nu =
1$ chiral edge mode, and an inner mode which propagates in the
opposite direction and has hole character, but is otherwise
similar to the chiral mode which occurs at the edge of a $\nu =
1/m$ QH system. (See Fig.~\ref{phconj}.) Long-range interactions
change the character of the collective modes.  In the limit of
strong coupling by long-range Coulomb interactions, the normal
modes that emerge are\cite{wen:adv:95,cllreview} a high-velocity
mode associated with fluctuations in the total electron charge
integrated perpendicularly to the edge, and a lower-velocity
mode associated with fluctuations in the distribution of a fixed
charge at a particular position along the edge.  The two modes 
propagate in opposite directions. The higher-velocity mode is
the microscopic realization of the EMP mode for a sharp $\nu =1
- 1/m$ edge. 

The occurrence of a counterpropagating mode with lower velocity
is, perhaps, counterintuitive. No such modes occur, for example,
in hydrodynamic theories of edge normal-mode structure.
Anticipation of a single lower-velocity long-lived
counterpropagating collective mode in the case of sharp edges is
grounded on fundamental notions of the microscopic theory of the
fractional QH effect, and on fundamental notions of the
phenomenology used to describe its edges. Experimental
verification for their existence would provide a powerful
confirmation of the predictive power of these theories. However,
time-domain studies\cite{ray:prb:92} of the propagation of edge
excitations at filling factor $\nu = 2/3$ have turned up no
evidence for this mode.

The main focus of the work presented here is to address 
properties of the sharp $\nu = 1-1/m$ edge, with the objective
of guiding future attempts to verify its normal-mode structure.
In Section~\ref{sec2} we discuss the excitation, propagation, and
detection of edge collective modes at a $\nu = 1 - 1/m$ edge.
The discussion in this section is phenomenological, and starts 
from the assumption that the edge charge is composed of 
contributions from two coupled chiral Luttinger liquids 
with opposite chirality. Such a model can be regarded as a
generalization of the Tomonaga-Luttinger (TL) model\cite{tlform}
that is used to describe conventional 1D systems like quantum
wires or 1D organic conductors. The parameters of the generalized
TL model Hamiltonian, which fix the velocities of the normal
modes and the way in which they are excited and detected,
are derived from a microscopic treatment of the underlying 
2D electron system. This calculation requires a careful
separation of long-range Coulomb and residual contributions to
the TL model parameters, explained in Section~\ref{sec3}.
The philosophy of this calculation is similar to that of
Landau-Silin theory\cite{silin} in which long-range Coulomb and
residual interactions between quasiparticles in metallic Fermi
liquids are carefully separated. We find that two
characteristics of the edge structure are most important in
determining the dispersions of the EMP mode and the
counterpropagating mode: the separation $d$ of the inner and
outer edges, and a velocity $v_{\text{J}}$ used to parameterize
the stiffness of the edge to neutral excitations. Evaluations of 
these parameters for a microscopic model of a sharp $\nu = 1 -
1/m$ edge are presented in in Sec.~\ref{sec4}. Numerical results
are given for the experimentally most relevant case $\nu = 2/3$.
We conclude in Section~\ref{sec5} with a discussion of the
implications of our results for possible experimental studies.
Some details of our calculations have been relegated to
Appendices.

\section{Edge wave packets at two-branch edges}
\label{sec2}
We have in previous work\cite{uz:prb:97} presented a detailed
theory of EMP wave-packet dynamics for single-branch
fractional-QH edges. Schemes for the excitation and detection 
of EMP wave packets were discussed, along with an analysis of 
the roles of noise and coupling to phonons of the host 
semiconductor. In this section we briefly present a
generalization of the most pertinent portions of this paper
to the case of present interest. 

We start from the assumption that the total electronic number
density integrated perpendicularly to the edge can be separated
into contributions from the inner and outer edges:  
$\varrho^{\text{(i)}}(x)$ and $\varrho^{\text{(o)}}(x)$.
Here $x$ is the 1D coordinate along the perimeter of the QH
sample, which we take to have length L. 
We write\cite{wen:int:92,convent}
\begin{mathletters}
\begin{eqnarray}
\varrho^{\text{(i)}}(x) &=& \sum_{q>0} \sqrt{\frac{q \,
\nu^{\text{(i)}}}{2\pi L}} \left( a^{\text{(i)}}_q\,
e^{i q x} + \big[a^{\text{(i)}}_q\big]^\dagger e^{- i
q x} \right) \,\, , \\
\varrho^{\text{(o)}}(x) &=& \sum_{q>0} \sqrt{\frac{q \,
\nu^{\text{(o)}}}{2\pi L}} \left( a^{\text{(o)}}_q\,
e^{- i q x} + \big[a^{\text{(o)}}_q\big]^\dagger e^{ i
q x} \right) \,\,\, .
\end{eqnarray}
\end{mathletters}
Here, $a^{\text{(i)}}_q$ ($a^{\text{(o)}}_q$) and $\big[
a^{\text{(i)}}_q \big]^\dagger$ ($\big[a^{\text{(o)}}_q
\big]^\dagger$) are Bose annihilation and creation operators for
chiral edge modes with 1D wave vector $q$ at the inner (outer)
edge. The values of the filling factors are $\nu^{\text{(i)}} :=
1 - \nu \equiv 1/m$ and $\nu^{\text{(o)}} := 1$. The commutation
relations implicit in the identification of creation and
annihilation operators follow, in the case of short-range
interactions, directly  from microscopic
considerations\cite{smg:prb:84,ahm:dbm:prb:85,jjp:prl:96,ahm:braz:96,ahm:prl:90}
which we elaborate on in further detail in Sec.~\ref{sec3}; see
also Appendix~\ref{append1}. The different sign of wave vector
associated with creation operators at the inner and outer edges
expresses the electron character of the outer chiral edge
excitations and the hole character of the inner chiral edge
excitations.\cite{ahm:prl:90}  

For general inter-particle interactions, we do not expect that
the low-energy effective TL Hamiltonian will be diagonal in the
boson fields associated with inner and outer edges. The normal
modes will be linear combinations of inner and outer edge modes
with coefficients which depend on the effective interactions
between inner and outer edges and vary from system to system.  
For the case of strong coupling due to long-range Coulomb
interactions, one of the normal modes is the EMP, and the other,
phonon-like, mode will have linear dispersion at long
wavelengths.\cite{wen:adv:95,mdjproc} The two sets of creation
and annihilation operators are related by a Bogoliubov
transformation:
\begin{equation}\label{bogoliub}
\left( \begin{array}{c} a^{\text{(o)}}_q \\ \big[
a^{\text{(i)}}_q\big]^\dagger \end{array} \right) = 
\left( \begin{array}{cc} {\mathrm cosh}\theta_q & -
{\mathrm sinh}\theta_q \\ -{\mathrm sinh}\theta_q &
{\mathrm cosh}\theta_q \end{array} \right) \left(
\begin{array}{c} \alpha^{\text{(pl)}}_q \\ \big[
\alpha^{\text{(ph)}}_q\big]^\dagger \end{array} \right) \,\, ,
\end{equation}
where the hyperbolic angle $\theta_q$ is, in general,
wave-vector dependent. When the Coulomb interaction is
unscreened, however, the coefficients become universal
at the longest length scales: ${\mathrm cosh}\theta_q\rightarrow
\sqrt{\nu^{\text{(o)}}/\nu}$ for $q\rightarrow 0$.

In the absence of an external perturbation, the diagonalized TL
model Hamiltonian $H_0$ of the edge is
\begin{equation}\label{unperturbed}
H_0 = \sum_{q>0} E^{\text{(pl)}}_q \, \big[
\alpha^{\text{(pl)}}_q\big]^\dagger \alpha^{\text{(pl)}}_q +
E^{\text{(ph)}}_q \, \big[\alpha^{\text{(ph)}}_q\big]^\dagger
\alpha^{\text{(ph)}}_q \quad ,
\end{equation}
with $E^{\text{(pl)}}_q$ and $E^{\text{(ph)}}_q$ denoting the
dispersion relations for the EMP and phonon modes, respectively.
In Sec.~\ref{sec3}, we derive explicit expressions for these
dispersion relations. Now suppose that an external time-dependent
potential $V^{\text{ext}}(t)$ couples electrostatically to the
edge. This can be achieved, e.g., by applying a voltage pulse to
a metallic gate close to the edge.\cite{ray:prb:92,uz:prb:97} In
general, the coupling of the inner and outer edges to the
external perturbation will differ:
\widetext
\top{-2.8cm}
\begin{mathletters}
\begin{eqnarray}
V^{\text{ext}}(t) &=& u(t) \int_0^L dx \,\, \left[
V^{\text{ext,o}} (x) \, \varrho^{\text{(o)}}(x) +
V^{\text{ext,i}}(x) \, \varrho^{\text{(i)}}(x) \right] \quad ,
\\ &=& u(t) \sum_{q>0} \sqrt{\frac{q L}{2\pi}} \left\{
\sqrt{\nu^{\text{(o)}}} \left( V^{\text{ext,o}}_{-q} \,
a^{\text{(o)}}_q + V^{\text{ext,o}}_{q} \, \big[
a^{\text{(o)}}_q\big]^\dagger \right) + \sqrt{\nu^{\text{(i)}}}
\left( V^{\text{ext,i}}_{q} \, a^{\text{(i)}}_q +
V^{\text{ext,i}}_{-q} \, \big[a^{\text{(i)}}_q\big]^\dagger
\right) \right\} \quad .
\end{eqnarray}
\end{mathletters}
\bottom{-2.7cm}
\narrowtext
\noindent
In this expression, the shape of the pulse is given by the
function $u(t)$, and geometrical details of the coupling between
the gate and the 1D edge densities at the inner and outer edges
are modeled by the functions $V^{\text{ext,i}}(x)$ and
$V^{\text{ext,o}}(x)$, respectively. The detailed form of these 
functions is determined by electrostatics. For practical
purposes, it is usually adequate to assume a {\em local-capacitor
model\/} where the metallic gate and the part of the edge located
in its immediate vicinity form the two `plates' of a
capacitor.\cite{uz:prb:97} In such a model, a capacitor which
covers the outer edge but not the inner edge would have 
$V^{\text{ext,i}}(x) = 0$. We will see that excitation of the
counterpropagating phonon mode requires differentiated coupling
to the inner and outer edges; the local-capacitor model suggests
that this could be achieved by arranging for an excitation gate
which covers only the outer part of the edge region. Alternately,
a side-gate geometry can also lead to stronger coupling to the
outer portion of the edge region.

Given the quadratic edge Hamiltonian, it is possible
to solve the time-dependent Schr\"odinger equation explicitly
for $H = H_0 + V^{\text{ext}}(t)$ with general pulse shape
$u(t)$ as detailed in Ref.~\onlinecite{uz:prb:97}. Wave packets
of edge modes can be engineered by appropriately adjusting the
characteristics of the voltage pulse.\cite{uz:prb:97} Wave
packets with narrow wave-vector distributions can be generated
by repeating a length-$T_{\text{exc}}$ pulse $N > 1$ times. 

One way to observe the time evolution of the charge disturbance
created by the external perturbation is to measure the charge
$Q(t)$ that is induced by evolving wave packets on metallic
gates situated close to the edge. In general, the gate will
respond differently to charge located at the inner and outer
edges:  
\widetext
\top{-2.8cm}
\begin{equation}
Q(t) = \int_0^L dx \,\, \left[ V^{\text{det,o}} (x-x_0) \,
\big< \varrho^{\text{(o)}}(x, t) \big> + V^{\text{det,i}}(x-x_0)
\, \big< \varrho^{\text{(i)}}(x, t) \big> \right] \quad ,
\end{equation}
where $x_0$ is the position (along the edge) of the observing
gate, and angle brackets $\langle \dots \rangle$ denote a
thermal average. The functions $V^{\text{det,i}}(x)$
[$V^{\text{det,o}}(x)$] model the coupling of the detecting gate
and the 2D electron system, which, we assume, can be
qualitatively understood using the local-capacitor model. An
explicit calculation following the formalism of
Ref.~\onlinecite{uz:prb:97} yields the result that there are two
contributions to the induced charge: $Q(t) = Q^{\text{(pl)}}(t)
+ Q^{\text{(ph)}}(t)$, corresponding to the EMP and phonon edge
wave packets:
%\widetext
%\top{-2.8cm}
\begin{equation}
Q^{\text{(pl,ph)}}(t) = 2 \text{Re} \left\{ \sum_{q>0}
Q^{\text{(pl,ph)}}_q (t) \exp{ [ i (\pm q x_0 - t
E_{q}^{\text{(pl,ph)}} / \hbar)
] } \right\} \, .
\end{equation}
In the small-$q$ limit, for unscreened Coulomb interactions, we
find that the Fourier components are given  by  
\begin{mathletters}\label{chargeres}
\begin{eqnarray}
Q^{\text{(pl)}}_q (t) &=& \frac{q L}{2\pi\nu} \, \big[
V^{\text{ext,o}}_{q} - (1-\nu) \, V^{\text{ext,i}}_{q} \big]
\big[ V^{\text{det,o}}_{-q} - (1-\nu) \, V^{\text{det,i}}_{-q}
\big] \,\, \frac{-i}{\hbar} \int_{-\infty}^{t} d\tau \,\,
u(\tau) \, \exp [ i \tau E_{q}^{\text{(pl)}} / \hbar ] \quad ,
\\ \label{neutralmode}
Q^{\text{(ph)}}_q (t) &=& \frac{q L}{2\pi\nu} \, (1 - \nu) \,
\big[ V^{\text{ext,o}}_{q} - V^{\text{ext,i}}_{q} \big] \big[
V^{\text{det,o}}_{-q} - V^{\text{det,i}}_{-q} \big] \,\,
\frac{-i}{\hbar} \int_{-\infty}^{t} d\tau \,\, u(\tau) \, \exp
[ i \tau E_{q}^{\text{(ph)}} / \hbar ] \quad .
\end{eqnarray}
\end{mathletters}
\bottom{-2.7cm}
\narrowtext
\noindent

{}From Eqs.~(\ref{chargeres}) we can deduce how the excitation
and detection of the two counterpropagating wave packets depends
on the device parameters. To create and observe the
{\em phonon\/} wave packet, both  exciting and observing gates
must couple differently to the inner and outer edges. This
condition probably requires that the metallic gates be positioned
with an accuracy better than $d$, the distance between the inner
and outer edges. The relative amplitudes of EMP and phonon wave
packets can be inferred from Eqs.~(\ref{chargeres}) as well: for
$V^{\text{ext,i}}(x)=V^{\text{det,i}}(x)=0$, e.g., the amplitude
of the phonon wave packet is smaller by a factor of $(1-\nu)$
than the amplitude of the EMP wave packet. This is probably
the largest relative amplitude which can be achieved. The group
velocities of the phonon and EMP wave packets will generally be
quite different;  
\begin{equation}
v^{\text{(pl,ph)}} = \frac{1}{\hbar}\,\,\frac{d\,
E_{q}^{\text{(pl,ph)}}}{d\, q} \Big|_{q =
{\tilde q}^{\text{(pl,ph)}}}\quad ,
\end{equation}
where ${\tilde q}^{\text{(pl,ph)}}$ is the median wave number
of the superposition of the modes forming the EMP and phonon
wave packets. (See Sec.~\ref{sec3} for explicit expressions for
the dispersion relations $E_{q}^{\text{(pl)}}$ and
$E_{q}^{\text{(ph)}}$. Typical values for the velocities are
given in Sec.~\ref{sec5}.) Since both wave packets are created
by the same external voltage-pulse characteristics, we
know\cite{uz:prb:97} that
$E_{{\tilde q}^{\text{(pl)}}}^{\text{(pl)}} = 
E_{{\tilde q}^{\text{(ph)}}}^{\text{(ph)}} = 2\pi \hbar /
T_{\text{exc}}$, which in turn allows us to predict that the
width (in real space) of the phonon wave packet is smaller by a
factor of the order of $v^{\text{(ph)}} / v^{\text{(pl)}}$ than
the width of the EMP wave packet.

\section{Separation of Contributions to the 1D Hamiltonian} 
\label{sec3}

In this Section, we develop a framework which reduces the 
task of determining generalized-TL-model parameters to 
a calculation of two microscopic quantities. The latter determine
the EMP and edge-phonon dispersion relations and the hyperbolic
mixing angle $\theta_q$ of Eq.~(\ref{bogoliub}).  

We consider a semi-infinite cylindrical QH sample which extends
from the edge near $y =0$ to $\infty$ in the $\hat y$ direction
and satisfies periodic boundary conditions in the $\hat x$
direction with $0\le x\le L$. This geometry is convenient for
calculations, and the results we obtain are readily applied to
experimentally realistic geometries. It is convenient to use the
Landau gauge for the single-particle basis states that describe
the motion of a 2D charged particle in a uniform transverse
magnetic field $B$. The Landau-gauge basis states factor into a
plane wave with 1D wave vector $k$, dependent on the
$x$-coordinate parallel to the edge, and a harmonic-oscillator
orbital of width $\ell$ centered at $-k\ell^2$ and dependent on
the $y$-coordinate perpendicular to the edge. Here $\ell
:=\sqrt{\hbar c/ |e B|}$ denotes the magnetic length. The
proportionality between the 1D wave vector parallel to the edge
and spatial displacement perpendicular to the edge, in
conjunction with the geometry of our QH sample, implies that, for
the many-particle ground state and its low-lying excitations, 
single-particle states with $k$ beyond a maximum value
$k_{\text{F0}}$ will be occupied with negligible probability.
It will be convenient for us to exploit this property by working 
in a truncated many-particle Fock space which includes only 
single-particle states with $k\le k_{\text{F0}}$. We choose the
zero for the $y$-coordinate such that a state with label $k$ has
its $y$-dependent orbital centered at $y=( k_{\text{F0}} - k )
\ell^2$. We use the simplest possible microscopic model which
will produce a sharp edge for the 2D electronic system by taking
the electrons to be confined by a coplanar neutralizing positive
background. To be specific, we take a background which would
exactly cancel the electron charge density if each electronic
orbital were occupied with probability $1 - 1/m$ out to the edge.
As we explain later, the electronic system is drawn in at the
edge, which permits us to let $k_{\text{F0}}$ coincide with the
edge of the positive  background.  

\begin{figure}[hbt]
\centerline{\epsfxsize3.35in\epsffile{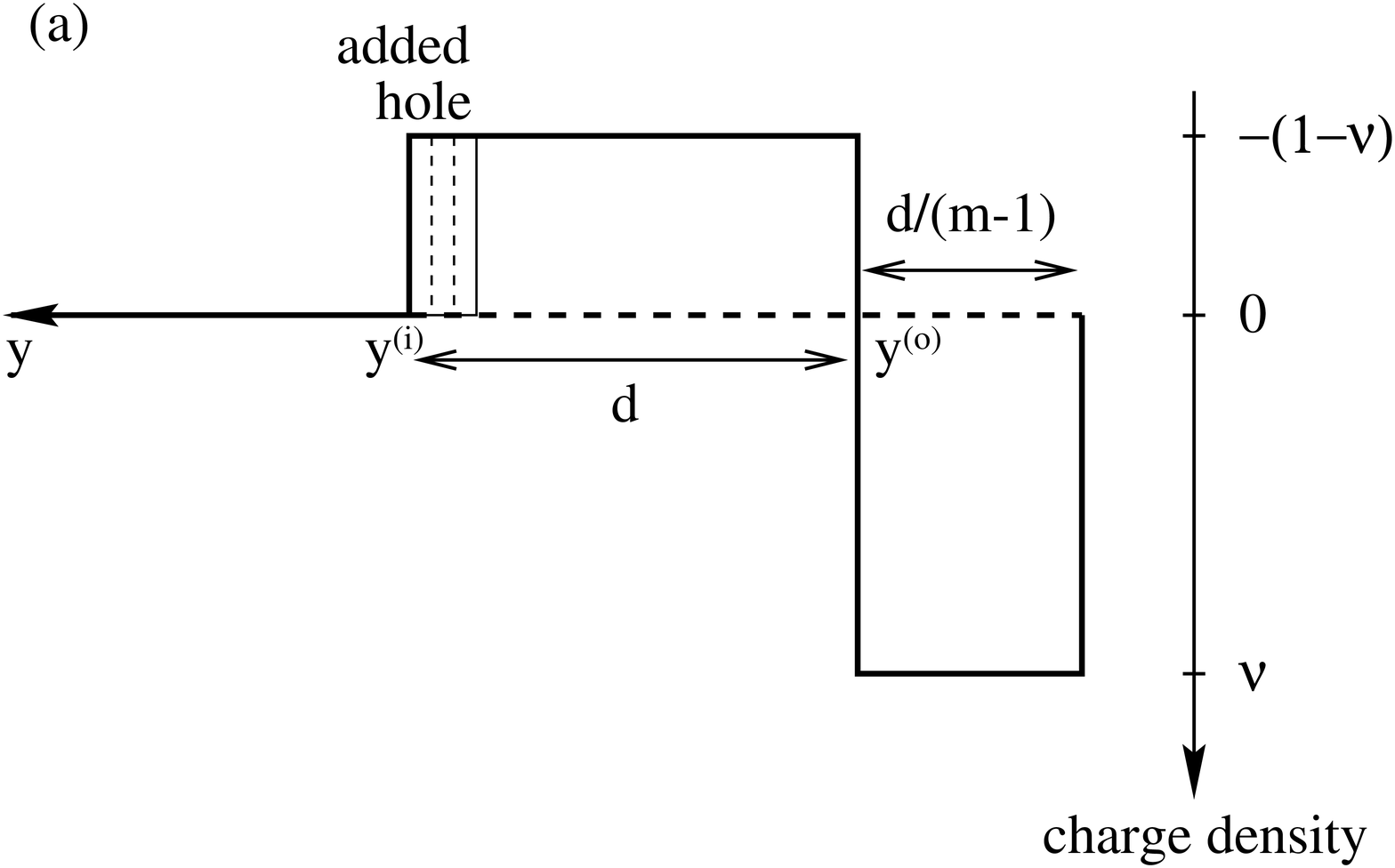} \hfill 
\epsfxsize3.35in \epsffile{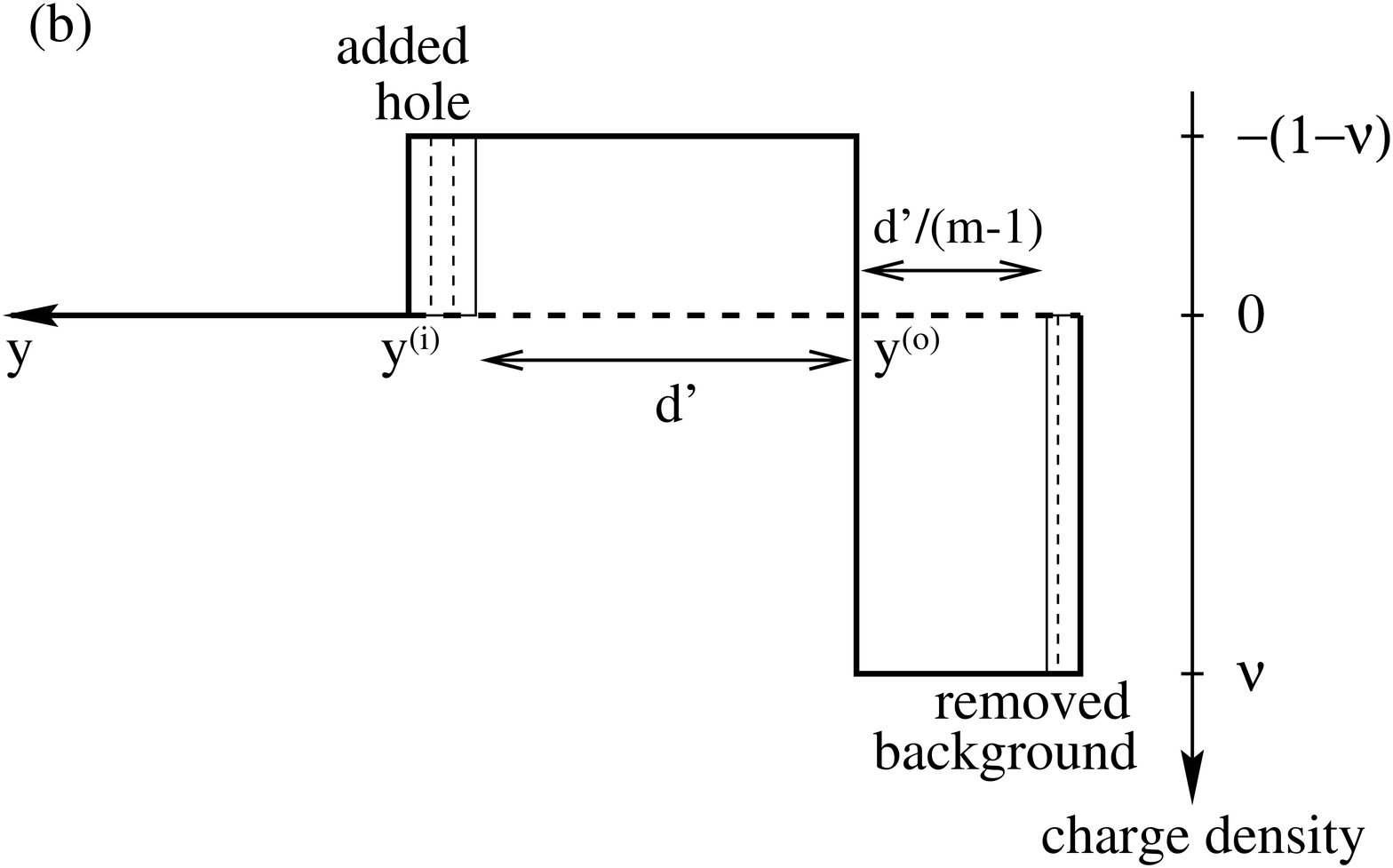}}
\vspace{0.2cm}
\caption{For a quantum-Hall system with bulk filling factor 
$\nu = 1 -  1/m$, the ground-state charge-density profile at a
sharp edge will have the dipolar strip that is illustrated
schematically in Fig.~\ref{phconj}. However, at any point along
the edge, the 1D charge density in the ground state, obtained by
integrating the 2D charge density along the coordinate
perpendicular to the edge, is zero. At long wave lengths,
parameters of the generalized Tomonaga-Luttinger model for the
edge are dominated by the long-range interactions between 1D
charge-density fluctuations. On the other hand, the
chiral-phonon-mode velocity is determined by smaller residual
interactions. In order to determine these accurately, we
introduce a fictitious model system in which the edge of the
background charge is adjusted to maintain zero 1D charge density
at each point along the edge. For example, when a hole is added
at the inner edge~[(a)], the background charge is reduced by
moving its outer edge as illustrated in~(b). Similarly, when the
outer edge is moved outward, the background edge is also moved
outward. (See text.) Here we measure densities in units of
$1/(2\pi\ell^2)$.}
\label{add}
\end{figure}

When edge effects are neglected, the many-particle Hamiltonian
truncated to the lowest Landau level is exactly particle-hole
symmetric. It follows that the ground state with $\nu=1-1/m$ is
precisely the particle-hole conjugate of the ground state at
$\nu = 1/m$.\cite{smg:prb:84,ahm:dbm:prb:85,ahmintro}
Particle-hole symmetry is broken at the edge of the system.
It has been conjectured\cite{ahm:prl:90} that the ground-state
electronic structure at the edge is formed by the particle-hole
conjugate of a $\nu = 1/m$ fractional-Hall state for
{\em holes\/} which is embedded into a filled-Landau-level state
for {\em electrons\/} which is truncated at $k_{\text{F0}}$.
For sharp edges, numerical studies support this
view.\cite{mdj:ahm:prl:91,kin:prb:92,dai:jpsj:93,greit:plb:94,yig:prl:94}
The calculations presented here provide further insight into 
the consistency of this scenario. In this paper, we find it
convenient to describe edge states of a $\nu=1-1/m$ QH sample
in the language of holes. The ground state then consists of
holes which have phase-separated into an inner strip, 
$y^{\text{(i)}} \le y < \infty$, which is in the incompressible 
state with filling factor $\nu^{\text{(i)}} = 1-\nu\equiv 1/m$,
and an outer strip with holes present for $0\equiv y^{\text{(b)}}
\le y \le y^{\text{(o)}}$ with hole filling factor
$\nu^{\text{(o)}} = 1$. For $y^{\text{(o)}} < y <y^{\text{(i)}}$,
no holes are present, i.e., the electron orbitals are
filled.\cite{realspace} Assuming overall charge neutrality,
$y^{\text{(o)}}$ and $y^{\text{(i)}}$ are not independent, and
the state is completely characterized by the separation $d$
between the inner and outer hole strips:
\begin{equation}
d \equiv y^{\text{(i)}} - y^{\text{(o)}} = \frac{\nu}{1-\nu}\,
\left[ y^{\text{(o)}} - y^{\text{(b)}} \right] \equiv
\frac{\nu}{1-\nu}\, y^{\text{(o)}} \quad .
\label{relate}
\end{equation}
For $d=0$, the outer strip is absent, and the system is
strictly neutral locally.  For $d>0$, the sample is still
{\em globally\/} neutral because of the presence of the uniform
neutralizing background, but a deviation from local neutrality
in form of a dipolar strip of charge exists; see
Figs.~\ref{phconj} and \ref{add}(a). This ground-state
configuration is still 1D-locally neutral, by which we mean
that, at any fixed position $x$ along the edge, the charge
density integrated perpendicularly to the edge yields zero.
Note that, in this hole language, charge fluctuations are
possible only at $y^{\text{(i)}}$ and $y^{\text{(o)}}$.
The outer edge of the outer hole strip at $y=y^{\text{(b)}}\equiv
0$ originates from the truncation of the Hilbert space in which
we perform the particle-hole conjugation and does {\em not\/}
support physical excitations.

Phenomenological\cite{wen:int:92} and
microscopic\cite{ahm:braz:96} considerations for the
non-interacting case have established that the excitations at a
chiral QH edge can be described as the excitations of a chiral
1D electron system. This is obvious for a filling factor equal
to one, because a filled Landau level is equivalent to a 1D
Fermi sea.\cite{uz:prb:96} But even more generally, for QH
systems with simple filling factors of the form $1/m$ with $m$
odd, the low-energy excited states are in one-to-one
correspondence to the low-lying states of a chiral 1D Fermi
gas.\cite{ahm:prl:93,ahm:braz:96} It can be expected that,
in the chiral system, the character of the low-lying excited
states remains unchanged in the presence of (even long-range)
interactions.\cite{stone:prb:92} As seen above, application of
particle-hole conjugation to describe the QH effect for systems
at filling factor $\nu=1-1/m$ leads to an edge-electronic
structure with two chiral edges; the inner edge at
$y^{\text{(i)}}$ (i.e., the outer edge of the hole system that is
in the $\nu^{\text{(i)}} = 1-\nu\equiv 1/m$ QH state) and the
outer edge at $y^{\text{(o)}}$ (i.e., the inner edge of the hole
system that is in the $\nu^{\text{(o)}} = 1$ filled-Landau-level
state). The validity of our description of the edge of such a QH
sample in terms of a generalized TL model rests on the assumption
that, even in the presence of long-range interactions, the
low-energy scattering processes conserve the number of particles
at the inner and outer edges separately.

Our calculation of the parameters of the generalized-TL
effective Hamiltonian is similar in spirit to the 
Landau-Silin theory\cite{silin} for charged Fermi liquids. In a
metal, interactions between quasiparticles, especially at small
scattering angles, can be totally dominated by the direct
Coulomb interaction. However, for some physical properties, e.g.,
the spin magnetic susceptibility, the Coulomb interaction cancels
out, leaving a dependence only on the weaker residual
interactions which reflect correlations between underlying
electronic degrees of freedom. Evaluation of the Fermi-liquid
parameters that determine the spin susceptibility requires that
the direct Coulomb interaction be carefully separated from
exchange and correlation contributions. Our main aim here is to 
estimate the phonon-mode velocity, which would vanish if only 
Coulomb interactions between 1D charge densities were included
in the generalized TL model Hamiltonian. In order to accurately
evaluate the important residual contributions to the effective
interactions in the model, we introduce an artificial model in
which long-range Coulomb interactions are eliminated by adjusting
the background charge to maintain 1D-local charge neutrality.
The energy, $\delta E$, of a state with a given 1D charge density
for the physical case of a fixed background charge differs from
the energy of the fictitious 1D-locally neutral system, $\delta
\tilde E$, because of the interactions between electrons and the
change in background, and because of the self-interaction energy
of the artificial change in background. For details, see
Appendix~\ref{append0}. Letting $\delta n^{\text{2D}}(\vec r)$
be the change relative to the ground state of the 2D electron
density and $\delta n^{\text{2D}}_{\text{bg}}(\vec r)$ be the
change in the background density necessary in the fictitious
1D-locally neutral system, we find that 
\begin{mathletters}\label{landausil}
\begin{equation}\label{allsilly}
\delta E = \delta\mu_{\text{bg}} + \delta \tilde E +
\delta E_{\text{C}} \quad ,
\end{equation}
with the definition
\begin{equation}\label{coulsil}
\delta E_{\text{C}} :=
\frac{e^2}{\epsilon} \int \frac{d^2 r\, d^2 r^\prime}{|
\vec r -\vec {r^\prime} |} \,\, \delta n^{\text{2D}}_{\text{bg}}
(\vec r) \left[ \delta n^{\text{2D}}(\vec {r^\prime} ) -
\frac{1}{2} \delta n^{\text{2D}}_{\text{bg}}(\vec {r^\prime} )
\right] \,\, ,
\end{equation}
\end{mathletters}
and a term $\delta\mu_{\text{bg}}$ contributing to the chemical
potential which is irrelevant for our considerations to follow
and will be dropped from now on. (See Appendix~\ref{append0}.)
Here, $\epsilon$ characterizes the dielectric environment of the
2D electron system.\cite{epsilon} The contribution $\delta\tilde
E[\delta n^{\text{2D}}(\vec r)]$ is the excitation energy in the
1D-locally neutral artificial system, and we will subsequently
refer to it as the neutral term; it contains only short-range
interaction contributions. The long-range Coulomb interaction is
contained in $\delta E_{\text{C}}$, the `Coulomb term'. In the
following subsections, we derive expressions for the two
corresponding contributions to the Tomonaga-Luttinger model
Hamiltonian which depend on two microscopic parameters
characterizing the edge. Section~\ref{sec4} describes the
evaluation of these parameters.  

\subsection{Edge-mode energies: Coulomb term}\label{coulcalc}
To evaluate the Coulomb term $\delta E_{\text{C}}$ for an edge
excitation, we have to find the 2D charge distributions $\delta
n^{\text{2D}}(\vec r)$ and $\delta n^{\text{2D}}_{\text{bg}}(
\vec r)$ that correspond to the 1D charge fluctuations
$\varrho^{\text{(i)}}(x)$ and $\varrho^{\text{(o)}}(x)$
associated with edge waves at inner and outer edges. Most
generally, we can write
\widetext
\top{-2.8cm}
\begin{mathletters}\label{2Dcharge}
\begin{eqnarray}
\delta n^{\text{2D}}(\vec r) &=& \varrho^{\text{(i)}}(x) \,\,
F^{\text{(i)}}\big(y -  y^{\text{(i)}} \big) +
\varrho^{\text{(o)}}(x)\,\, F^{\text{(o)}}\big(y - y^{\text{(o)}}
\big) \quad , \\ \delta n^{\text{2D}}_{\text{bg}}(\vec r) &=& 
\left[ \varrho^{\text{(i)}}(x) + \varrho^{\text{(o)}}(x) \right]
\,\, F^{\text{(b)}}\big(y - y^{\text{(b)}}\big) \quad .
\end{eqnarray}
\end{mathletters}
The structure of the transverse density profile at the inner and
outer edges as well as at the physical boundary of the sample
enters through the form factors $F^{\text{(i)}}(y)$,
$F^{\text{(o)}}(y)$, and $F^{\text{(b)}}(y)$, respectively. Using
the Fourier representation, and defining the coupling functions
\beq\label{couplfunct}
F_q^{(r s)} := \frac{e^2}{\epsilon} \, \, 2\, \int dy \,\,
dy^\prime \,\,\,\,{\mathrm K_0}(q \, |y - y^\prime|) \,\, F^{(r)}
\big( y - y^{(r)} \big) \, F^{(s)}\big(y^\prime - y^{(s)}\big)
\quad ,
\eeq
where the indices $r, s\in \{\text{i,o,b}\}$ and ${\mathrm K_0}$
denotes a modified Bessel function of zeroth order, we express
the Coulomb term in a form which will be convenient for
identifying its contribution to the TL model\cite{tlform}
Hamiltonian:
\begin{equation}\label{coulterm}
\delta E_{\text{C}} = \frac{2\pi\hbar}{L} \,\sum_{q>0} \lambda_q
\, \big[ \varrho^{\text{(i)}}_q + \varrho^{\text{(o)}}_q \big]
\big[ \varrho^{\text{(i)}}_{-q} + \varrho^{\text{(o)}}_{-q}\big]
+ \frac{2\pi\hbar}{L} \,\sum_{q>0} v_{\text{C}} \,\big[
\varrho^{\text{(o)}}_q \varrho^{\text{(o)}}_{-q} -
\varrho^{\text{(i)}}_q \varrho^{\text{(i)}}_{-q} \big] \quad .
\end{equation}
\bottom{-2.7cm}
\narrowtext
\noindent
The parameters $\lambda_q$ and $v_{\text{C}}$ have the units of
velocity and are given by
\begin{mathletters}\label{lambdavc}
\begin{eqnarray}\label{plasvel1}
2\pi\hbar\,\,\lambda_q &=& F_q^{\text{(bi)}} + F_q^{\text{(bo)}}
- F_q^{\text{(bb)}} \quad , \\
2\pi\hbar\,\, v_{\text{C}} &=& F_q^{\text{(bo)}} -
F_q^{\text{(bi)}} \quad .
\end{eqnarray}
\end{mathletters}
(In most cases, the wave-vector dependence of $v_{\text{C}}$ will
be unimportant.) In Eq.~(\ref{coulterm}), we have separated
$\delta E_{\text{C}}$ into a term dependent only on the total 1D
charge fluctuation at the edge and a term which occurs because of
the spatial separation of inner and outer edges. The first term
in Eq.~(\ref{coulterm}) corresponds to the familiar\cite{emp} 
EMP mode, which becomes one of the edge normal modes if
long-range Coulomb interaction is
present\cite{wen:adv:95,mdjproc} (see also Sec.~\ref{normsec}
below). In that case, the second term in Eq.~(\ref{coulterm})
which involves the velocity $v_{\text{C}}$ becomes important only
at large wave vectors. Note that, if $\delta E_{\text{C}}$ were
the only contribution to the edge excitation energy, the
counterpropagating phonon mode would have zero velocity; see
Sec.~\ref{normsec} below. In the small-$q$ limit,
Eq.~(\ref{couplfunct}) simplifies to
\beq
F_q^{(r s)} = - \frac{e^2}{\epsilon} \,\, 2\, \left[ \ln\left(
4\alpha^2\, q\,\ell\right) + \Delta^{(r s)} \right] \quad ,
\eeq
where $\alpha = \sqrt{e^C / 8} \approx 0.47$ with $C=0.577\dots$
being Euler's constant, and
\widetext
\top{-2.8cm}
\beq\label{residude}
\Delta^{(r s)} = \int d y \,\, dy^\prime \,\,\,\, \ln \left(
\big| y - y^\prime + y^{(r)} - y^{(s)} \big| / \ell \right)\,
\,\, F^{(r)}(y)\,\, F^{(s)}(y^\prime) \quad .
\eeq
\bottom{-2.7cm}
\narrowtext
\noindent
[Some analytical details of the function $\Delta^{(r s)}$ are
known\cite{uz:prb:96} for the special case of $F^{(r)}(y) =
F^{(s)}(y) = \exp(-y^2/ \ell^2) /(\sqrt{\pi}\ell)$.] In general,
Eqs.~(\ref{lambdavc}) specialize in the small-$q$ limit to
\begin{mathletters}\label{somecoul}
\begin{eqnarray}\label{plasvel2}
\lambda_q &=& - \frac{e^2}{\epsilon\hbar} \, \frac{1}{\pi} \,\,
\ln \left(16\alpha^3 \, \Upsilon_\lambda(d)\,\frac{1-\nu}{\nu^2}
\,\frac{d^2}{\ell}\,\, q \right) \quad , \\
v_{\text{C}} &=& - \frac{e^2}{\epsilon\hbar}\, \frac{1}{\pi}\,\,
\ln \left[ \Upsilon_{\text{C}}(d)\,(1-\nu)\right] \quad ,
\end{eqnarray}
\end{mathletters}
The fact that $\lambda_q \to \infty$ for $q \to 0$ results from
the long range of the Coulomb interaction. The $d$-dependent
factors $\Upsilon_\lambda(d)$ and $\Upsilon_{\text{C}}(d)$
account for the details of the transverse density profile. Both
approach unity for $d\gtrsim\ell$. The microscopic parameter $d$
must be determined to fix the TL model parameters. Its value for
filling factor $\nu = 2/3$ is calculated in Sec.~\ref{sec4},
where we find $d_{\text{2/3}} \approx 1.7\, \ell$. This value is
consistent with numerical studies\cite{yig:prl:94} performed for
systems with up to 50 electrons. We determined the correction
factors $\Upsilon_\lambda(d_{\text{2/3}}) \approx 
\Upsilon_{\text{C}}(d_{\text{2/3}}) \approx 0.86$. See
Appendix~\ref{transedge} for that calculation and a detailed
discussion of the transverse density profile for edge
excitations. Our result [Eq.~(\ref{plasvel2})] for the EMP
dispersion relation is similar to the one obtained in
hydrodynamic theories\cite{emp} if we interpret $d^2/\ell$ as the
effective width of the edge region.

\subsection{Edge-mode energies: neutral term}
We now evaluate the neutral term $\delta\tilde E$ in
Eq.~(\ref{allsilly}). This is the energy of an edge excitation
in a fictitious system where the neutralizing background is
adjusted so that the charge density integrated perpendicularly
to the edge vanishes at any fixed point along the edge. We call
this property `1D-local neutrality'.  With excitations present,
the inner and outer edges move to new positions $\tilde
y^{\text{(i)}},\tilde y^{\text{(o)}}$, with a changed separation
$D=\tilde y^{\text{(i)}}-\tilde y^{\text{(o)}}$. In the
fictitious system where 1D-local neutrality is maintained, the
background charge ends not at $y^{\text{(b)}}=0$ but instead at
some new position $\tilde y^{\text{(b)}}$. When the density of
holes varies with $x$, all of $\tilde y^{\text{(i)}},\,
\tilde y^{\text{(o)}},\, \tilde y^{\text{(b)}},\, D$ will also
depend on $x$. Requiring 1D-local charge neutrality at each
position $x$ along the edge yields $D(x) = (m-1) [\tilde
y^{\text{(o)}}(x) - \tilde y^{\text{(b)}}(x)]$, exactly like
Eq.~(\ref{relate}). The neutral-edge system is completely
characterized by $D(x)$, and the energy $\delta\tilde E$ can be
expressed as a functional of $D(x)$, or, more conveniently, as a
functional of $[D(x) - d]$ where $d$ is the ground-state
separation of the inner and outer edges. In order to quantize
this energy functional, we must express $D(x)$ in terms of the
charge-density contributions from inner and outer edges. The
relation between the deviation of $D(x)$ from its ground-state
value $d$ and the 1D charge fluctuations localized at the inner
and outer edges can be derived straightforwardly; it is
\begin{equation}\label{substitute}
D(x) - d = - 2\pi \ell^2 \left[ \varrho^{\text{(o)}}(x) +
\frac{1}{1-\nu}\, \varrho^{\text{(i)}}(x) \right] \quad .
\end{equation}
Equation~(\ref{substitute}) is an exact statement and follows
from the fact that edge waves at the inner (outer) edge
correspond to rigid deformations of the 2D ground-state density
profile for the inner (outer) QH strip. See
Appendix~\ref{transedge} for details. When $\varrho^{\text{(i)}}
(x) = - \varrho^{\text{(o)}}(x)/m $, both edges suffer identical
displacements and the distance between them is not altered.  

As the configuration with $D(x)\equiv d$ is the ground state,
the zeroth- and first-order terms in the functional expansion of
$\delta\tilde E$ with respect to\ $[D(x)-d]$ vanish. Unlike the
Coulomb term, this contribution to the energy will be local for
long-wavelength excitations, allowing us to parameterize the
coefficient of the quadratic term in terms of a single parameter,
$v_{\text{J}}$, with units of velocity:
\begin{equation}\label{quadrat}
\delta\tilde E = \frac{v_{\text{J}}}{2}\, \frac{1- \nu}{\nu} \,
\frac{\hbar}{2\pi\ell^2} \, \int d x \,\,
\left[\frac{D(x) - d}{\ell}\right]^2 \quad .
\end{equation}
Expressing the distance between inner and outer edges in terms of
inner and outer edge charge densities using
Eq.~(\ref{substitute}), and Fourier transforming allows us to
write the short-range term in a convenient TL form:\cite{tlform}
\widetext
\top{-2.8cm}
\begin{equation}\label{neutrterm}
\delta\tilde E = \frac{2\pi\hbar}{L}\,\, v_{\text{J}}\,\,
\sum_{q>0} \left\{ \frac{1-\nu}{\nu} \,\, \varrho^{\text{(o)}}_q
\varrho^{\text{(o)}}_{-q} + \frac{1}{\nu (1-\nu)} \,\,
\varrho^{\text{(i)}}_q \varrho^{\text{(i)}}_{-q} + \frac{1}{\nu}
\,\,\big[ \varrho^{\text{(o)}}_q \varrho^{\text{(i)}}_{-q} +
\varrho^{\text{(i)}}_q \varrho^{\text{(o)}}_{-q} \big] \right\}
\quad .
\end{equation}
\bottom{-2.7cm}
\narrowtext
\noindent
We show in Sec.~\ref{sec4} how to determine the velocity
$v_{\text{J}}$. An analytical result (valid for $d\gg\ell$) is
\begin{equation}\label{uliconst}
v_{\text{J}} = -\frac{e^2}{\epsilon\hbar}\,\frac{1}{\pi}\,\left[
\nu\,\ln(\nu) + (1-\nu)\,\ln(1-\nu) \right] \quad.
\end{equation}
Our calculation (outlined in Sec.~\ref{sec4} and detailed in
Appendix~\ref{append1}) shows, however, that the ground-state
separation $d_{\text{2/3}}$ of the inner and outer edges for the
case of $\nu = 2/3$ is not particularly large, so corrections to
the asymptotic formula [Eq.~(\ref{uliconst})] have to be taken
into account. As an improved result for $\nu = 2/3$, we find
$v_{\text{J}} \approx 0.24\, e^2/(\epsilon\hbar)$.

\subsection{Dispersion of EMP normal modes}\label{normsec}
The low-energy, small-wave-vector effective 1D Hamiltonian for
the edge at filling factor $\nu = 1-1/m$ can be written in form
of a TL Hamiltonian;\cite{tlform} it is given by
\begin{equation}\label{quantize}
\delta E_{\text{C}} + \delta\tilde E \to H_{\text{TL}} \quad ,
\end{equation}
with the terms $\delta E_{\text{C}}$ and $\delta\tilde E$ taken
from Eqs.~(\ref{coulterm}) and (\ref{neutrterm}), respectively.
Equation~(\ref{quantize}) signifies that we obtain the TL
Hamiltonian from our energy calculations by considering
the 1D density fluctuations as operators which
have the appropriate chiral-Luttinger-liquid commutation
relations.\cite{wen:int:92} A straightforward Bogoliubov
transformation\cite{wen:adv:95} [Eq.~(\ref{bogoliub})] to the
normal modes yields the diagonal Hamiltonian of
Eq.~(\ref{unperturbed}). In the small-wave-vector limit (where
$\lambda_q \gg v_{\text{J}}, v_{\text{C}}$), we find for the
dispersions of the EMP and phonon normal modes
\begin{mathletters}\label{energies}
\begin{eqnarray}
E^{\text{(pl)}}_q &=& \hbar q \,\, \nu  \lambda_q \quad , \\
E^{\text{(ph)}}_q &=& \hbar q \,\, v_{\text{J}} \quad .
\end{eqnarray}
\end{mathletters}
[The expression for $\lambda_q$ in its most general form is
given in Eq.~(\ref{plasvel1}). With our approximations used, we
find Eq.~(\ref{plasvel2}).] We see that the energy of the EMP
normal mode is due primarily to the Coulomb interaction; the
separation $d$ of the inner and outer edges in the ground state
enters prominently because it determines the effective width of
the edge region. The energy of the phonon-like mode, however, is
naturally given by the velocity $v_{\text{J}}$, because that
quantity measures the energy of excitations that preserve
1D-local neutrality in the system.

\section{Evaluation of edge width and phonon-mode velocity} 
\label{sec4}

We have shown that the $\nu = 1 - 1/m$ sharp-edge Hamiltonian
can be expressed in terms of two characteristic parameters: the
ground-state separation $d$ between inner and outer edges, and
the velocity $v_{\text{J}}$. In this Section, we determine both
quantities simultaneously by calculating the energy change due
to a hole transfer from the inner to outer incompressible strips
at a neutral edge.

\begin{figure}[hbt]
\epsfxsize3.3in
\centerline{\epsffile{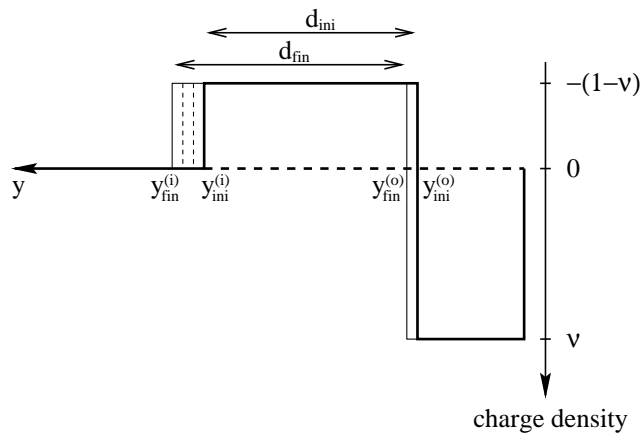}}
\vspace{0.2cm}
\caption{The process we consider to calculate the parameters $d$
and $v_{\text{J}}$. We consider configurations in which the holes
have phase-separated into two incompressible strips as in the 
ground state at filling factor $\nu = 1-1/m$, but allow the
separation $y^{\text{(i)}}_{\text{ini}} -
y^{\text{(o)}}_{\text{ini}}$ of the two strips to differ from
$d$ ($\equiv$~the strip separation in the ground state). We
imagine then that a hole is transferred from the edge of the
inner strip to the edge of the outer strip, preserving the
1D-local neutrality of the reference state. No adjustment of the
background is necessary, and the overall shape of the 1D density
profile is similar to its shape in the ground state. We are able
to extract the values of $d$ and $v_{\text{J}}$ from the energy
change produced by the hole transfer.}
\label{transfer}
\end{figure}

Consider a configuration that differs from the ground state only
by the transfers of an arbitrary number of holes between inner
and outer strips. Such a state is 1D-locally neutral, and its
charge profile perpendicular to the edge looks similar to that
of the ground state. However, we allow the separation of the
inner and outer edges ($\equiv y^{\text{(i)}}_{\text{ini}} -
y^{\text{(o)}}_{\text{ini}}$) to differ from the value $d$ for
the ground state. (See Fig.~\ref{transfer}.) The energy of such
an excited state is given by $\delta E_{\text{ini}}$ ($\equiv
\delta\tilde E_{\text{ini}}$ because no adjustment of the
background is necessary to ensure 1D-local neutrality).
If $y^{\text{(i)}}_{\text{ini}} - y^{\text{(o)}}_{\text{ini}}$
is not too much different from $d$, we can write
\begin{equation}
\delta E_{\text{ini}} \equiv \delta \tilde E_{\text{ini}} =
\frac{v_{\text{J}}}{2}\, \frac{1-\nu}{\nu}\, \frac{\hbar L}{2\pi
\ell^2}\,\left[ \frac{\big[y^{\text{(i)}}_{\text{ini}} -
y^{\text{(o)}}_{\text{ini}}\big] - d}{\ell} \right]^2 \quad ,
\end{equation}
which is a specialization of Eq.~(\ref{quadrat}) to the case of
an excitation with a transverse density profile that is uniform
along the edge.

Now we transfer one extra hole from the inner edge to the outer
one (see Fig.~\ref{transfer}). This changes the separation of
the two edges by
\begin{equation}
\Delta D(x) = \frac{\nu}{1-\nu}\, \frac{2\pi\ell^2}{L} \quad .
\end{equation}
For the corresponding energy change, we find
\begin{equation}\label{transenerg}
\Delta(\delta E_{\text{ini}}) \approx \frac{\hbar
v_{\text{J}}}{\ell}\, \frac{\big[ y^{\text{(i)}}_{\text{ini}} -
y^{\text{(o)}}_{\text{ini}}\big] - d}{\ell} \quad ,
\end{equation}
where we neglected a term that is small if the relation
$$ \frac{\big[ y^{\text{(i)}}_{\text{ini}} -
y^{\text{(o)}}_{\text{ini}}\big] - d}{\ell} \gg 2\pi\,
\frac{\nu}{1-\nu}\, \frac{\ell}{L} $$
holds. As the perimeter $L$ of the edge in typical QH samples is
usually many magnetic lengths, such an assumption is valid
except for an extremely narrow interval around the point
$y^{\text{(i)}}_{\text{ini}} - y^{\text{(o)}}_{\text{ini}} = d$.

\begin{figure}[hbt]
\epsfxsize3.5in
\centerline{\epsffile{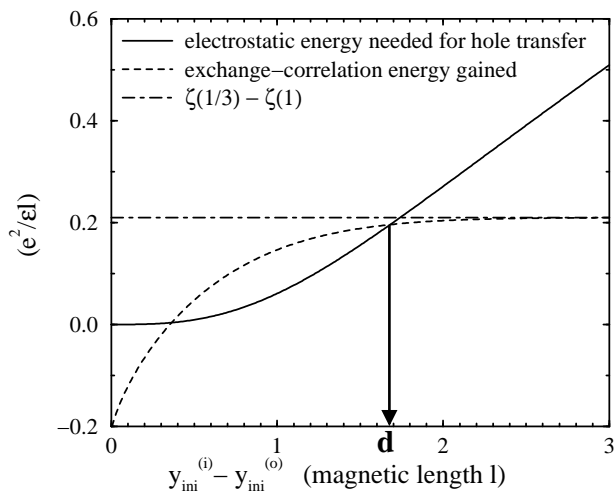}}
\vspace{0.2cm}
\caption{Energy balance for the transfer of a hole from the
inner edge to the outer one (see Fig.~\ref{transfer}). The
curves are calculated for a filling factor $\nu = 2/3$. The
electrostatic energy required to transfer the hole (solid curve)
is the work performed against the external potential stemming
from the dipolar strip of charge; see Fig.~\ref{phconj}. Details
of its evaluation are given in the appendix. This portion of the
energy is linear in the separation $y^{\text{(i)}}_{\text{ini}}
- y^{\text{(o)}}_{\text{ini}}$ of the inner and outer edges for
separations larger than the magnetic length. The
exchange-correlation energy gain is given approximately by
$|\zeta(1)| - |\zeta(1/3)|$. Corrections to the simple expression
for the exchange-correlation energy gain detailed in the
appendix become important at smaller inter-edge distance. The
full result for the exchange-correlation energy gain is given by
the dashed curve. The point where the two curves cross gives the
equilibrium edge separation within our variational two-strip
model. The approximations used in our calculation of the
exchange-correlation energy are not valid for strip separations
much smaller than the magnetic length $\ell$, and the crossing
of the curves at the smaller value of
$y^{\text{(i)}}_{\text{ini}} - y^{\text{(o)}}_{\text{ini}}$ is
unphysical. The other crossing occurs in a regime where our
approximations apply. From the point of crossing we conclude
that $d\approx 1.7\,\ell$. At this value of $d$, the
exchange-correlation energy gain is nearly constant and the 
electrostatic-energy-cost curve is nearly linear. From its slope
we obtain $v_{\text{J}}\approx 0.24\, e^2/(\epsilon\hbar)$.}
\label{numsep}
\end{figure}

To determine the parameters $d$ and $v_{\text{J}}$, we performed
a microscopic calculation of the energy on the left-hand-side of
Eq.~(\ref{transenerg}).  This turns out to indeed yield an
expression of the form of the right-hand-side, with suitable
choices of the parameters $d$ and $v_{\text{J}}$. The
equilibrium separation between inner and outer edges is reached
when the energy change associated with hole transfer vanishes.
A summary of the calculation details is relegated to
Appendix~\ref{append1}. Here we explain the main ingredients and
report numerical results for filling factor $\nu = 2/3$, which
are summarized in Fig.~\ref{numsep}.

The energy required to perform the transfer of a hole from the
inner edge to the outer one has several contributions. Some are
conveniently expressed in terms of $\zeta(\gamma)$, the energy
per particle in a homogeneous QH state of filling factor
$\gamma$ in the presence of a uniform coplanar neutralizing
background.\cite{perp} Hartree and exchange-correlation
contributions to the energy change are treated separately in the
calculation. The essence of the energetics at the edge can be
understood by the following simple argument. First we remove a
hole from the edge of the inner strip which is in a fractional-QH
state of filling factor $1-\nu$. The loss of exchange-correlation
energy is $|\zeta(1-\nu)| $. Adding this hole to the edge of the
outer strip gives a gain in exchange-correlation energy which is
close to $ |\zeta(1)| $, provided that the width of the outer
strip is larger than the magnetic length. (Since the outer strip
is a simple filled-Landau-level state, it is easy to incorporate
finite-thickness corrections to its addition energy, and we do
so as detailed in the appendix.) Since $|\zeta(1-\nu)| <
|\zeta(1)| $, there is a net gain in {\em exchange-correlation\/}
energy when transferring holes from the inner strip to the outer
one in that situation. This gain is balanced by the increase in
{\em electrostatic\/} energy that comes about due to the
existence of the dipolar strip of charge; see  Fig.~\ref{phconj}.
The hole that is being transferred is brought closer to the outer
part of the dipolar strip which electrostatically repels holes.
The separation of the two edges in the state where the gain in
exchange-correlation energy for the hole transfer is exactly
off-set by the loss in electrostatic energy is the ground-state
separation $d$. The electrostatic energy cost of hole transfer
increases linearly with $d$ for $d > \ell $. Comparing with
Eq.~(\ref{transenerg}), we see that, in this approximation, the
slope of the curve for the electrostatic contribution to the
transfer energy is $\hbar v_{\text{J}} / \ell$. This simple
picture requires a number of modifications which are detailed in
the appendix but, as illustrated for $\nu = 2/3$ in
Fig.~\ref{numsep}, these have little quantitative importance. 

\section{Discussion of Experimental Implications}
\label{sec5}

We have determined the conditions under which it is possible to
excite and observe two counterpropagating EMP wave packets at
the edge of a QH sample that is at filling factor $\nu =1-1/m$.
It is important that the geometry of the sample allows for an
external potential that is different at the positions of the
inner and outer edges. According to the calculation of the
previous section, the separation $d$ of the two strips for
filling factor $\nu = 2/3$ is $d_{\text{2/3}}\approx 1.7\,\ell$.
In typical magnetic fields, this corresponds to $d_{\text{2/3}}
\sim 20$~nm. For a top gate, significant differential coupling
to inner and outer edges would require that the distance to the
gate not be too much larger than $\sim 20$~nm and that its edge
be positioned relative to the QH edge with an accuracy of better
than $\sim 20$~nm. Both these conditions appear to be realizable.

The result we have obtained for the EMP wave-packet group
velocity is 
\begin{equation}\label{concluemp}
v^{\text{(pl)}} = - \nu\,\frac{e^2}{\epsilon\hbar}\,
\frac{1}{\pi} \,\, \left[ \ln \left(16\alpha^3 \,\Upsilon_\lambda
(d)\, \frac{1-\nu}{\nu^2}\, \frac{d^2}{\ell} \,\,
{\tilde q}^{\text{(pl)}} \right) + 1 \right] \, ,
\end{equation}
where ${\tilde q}^{\text{(pl)}}$ is the characteristic wave
vector of the dominant charge fluctuation in this wave packet.
Specializing Eq.~(\ref{concluemp}) to the case of the dielectric 
environment of typical 2D electron systems in GaAs,\cite{epsilon}
taking a QH sample with $\nu = 2/3$, and assuming
${\tilde q}^{\text{(pl)}} L  \ll L/ \ell$, we find that  
$v^{\text{(pl)}}_{\text{2/3}}\sim 70 \times \ln[L/ (50\ell)]
\,\, {\mathrm \mu m/ns}$. The phonon wave packet moves in a
direction opposite to that of the EMP wave packet, and has
linear dispersion with velocity $v^{\text{(ph)}} \equiv
v_{\text{J}}$. For $\nu = 2/3$, we have found that $v_{\text{J}}
\approx 0.24\,e^2/(\epsilon\hbar)$. In typical samples, we
therefore have $v^{\text{(ph)}}_{\text{2/3}}\sim 70 \,\,
{\mathrm \mu m/ns}$. The ratio $v^{\text{(pl)}}_{\text{2/3}}/
v^{\text{(ph)}}_{\text{2/3}}$ turns out to be of the order of
$\ln[L/ (50\ell)]$; this number is $\sim 8$ for the experiment
reported in Ref.~\onlinecite{ray:prb:92}. The relative width
of the two wave packets is inversely proportional to the ratio
of their respective velocities; the phonon wave packet will
therefore be much more narrow (in its 1D extension along the
edge) because it is much slower than the EMP wave packet.
We expect the numerical group-velocity estimates given here to
be realistic for the case of a sharp edge with an external
potential sufficiently similar to that produced by the coplanar
neutralizing charge used in these microscopic calculations.  
It appears likely to us that sharp edges will occur only in
specially prepared QH samples, for example in those prepared
using a cleaved-edge overgrowth technique.\cite{smooth} We remark
that this technique appears to be compatible with side-gate-based
capacitive coupling which we believe will produce the
differentiation necessary to excite the phonon modes.
The microscopic formalism developed in this work can, in
principle, be elaborated to model the details of a specific
sample and arrive at precise predictions for the relative
velocities of the two modes. The microscopic electronic structure
at smooth edges is presently not well
understood,\cite{uz:ahm:to_come} even for the simpler case where
the bulk filling factor is an integer. Nevertheless, it appears
clear that, for very smooth edges, 1D electron-gas models are not
appropriate. The excitation spectrum will have many collective
modes,\cite{glaz:prl:94} and each of these will, in general,
decay into incoherent particle-hole excitations at a finite rate.
If a sample with a sharp edge can be fabricated, the present
calculations suggest that group velocities of the modes are slow
enough to permit the use of capacitive coupling to detect
wave-packet evolution, and fast enough to permit several orbits
around a macroscopic sample to occur before the wave packet is
dissipated through its coupling to bulk phonon modes of the host
semiconductor.\cite{uz:prb:97}

\acknowledgements{
It is a pleasure to thank R.~C.~Ashoori, S.~Conti, G.~Ernst,
M.~R.~Geller, K.~v.~Klitzing, W.~L.~Schaich, and G.~Vignale for
stimulating discussions. This work was funded in part by NSF
Grant Nos.~DMR-9714055 (Indiana) and DMR-9632141 (Florida). U.Z.\
is partially supported by Stu\-dien\-stif\-tung des deutschen
Volkes (Bonn, Germany).
}

\widetext
\pagebreak

\begin{appendix}

\narrowtext

\section{Landau-Silin-type separation of
Coulomb and short-range interactions} \label{append0}

In this section, we show briefly how the separation of the
Coulomb and short-range pieces of the interaction leads to
Eqs.~(\ref{landausil}).

We start from the ground state of an edge which has
a density profile as depicted schematically in Fig.~\ref{phconj}.
Our goal is to find the energy $\delta E$ it costs to make an
excitation that leads to a deviation $\delta n^{\text{2D}}(\vec
r )$ from the ground-state density profile. To separate
long-range and short-range contributions to $\delta E$, we relate
our physical system to a fictitious system which has only
short-range forces, because any excitation $\delta n^{\text{2D}}
(\vec r )$ is simultaneously followed by an adjustment of the
background charge density $\delta n^{\text{2D}}_{\text{bg}}(\vec
r )$ that restores 1D-local neutrality. Obviously, the amount of
energy $\delta\tilde E$ that it takes to make an excitation
$\delta n^{\text{2D}}(\vec r )$ in the fictitious 1D-locally
neutral system differs from $\delta E$ by the energy necessary
for adjusting the background charge:
\widetext
\top{-2.8cm}
\begin{equation}\label{separcon}
\delta\tilde E = \delta E + \frac{e^2}{\epsilon} \int d^2 r
\,\, d^2 r^\prime \,\,\, \frac{1}{|\vec r - \vec {r^\prime}|}
\,\, \delta n^{\text{2D}}_{\text{bg}}(\vec r) \,\, \left\{
\frac{1}{2} \delta n^{\text{2D}}_{\text{bg}}(\vec {r^\prime} ) +
n^{\text{2D}}_{\text{bg}}(\vec {r^\prime} ) - \delta
n^{\text{2D}}(\vec {r^\prime} ) - n^{\text{2D}}(\vec {r^\prime} )
\right\}\quad .
\end{equation}
\bottom{-2.7cm}
\narrowtext
\noindent
The first term in the curly brackets of Eq.~(\ref{separcon})
comes from the self-interaction of the adjusted piece of the
background, the second term is the interaction energy of the
adjusted background piece with the ground-state background-charge
distribution denoted by $n^{\text{2D}}_{\text{bg}}(\vec r)$, the
third one is the interaction energy of the charged electronic
excitation with the adjusted piece of the background, and the
last term comes from the interaction of the electronic
ground-state charge distribution $n^{\text{2D}}(\vec r)$ with
the adjusted background piece. We arrive readily at
Eqs.~(\ref{landausil}) if we define
\begin{equation}
\delta\mu_{\text{bg}} := \frac{e^2}{\epsilon} \int \frac{d^2 r
\,\, d^2 r^\prime}{|\vec r - \vec {r^\prime}|} \, \delta
n^{\text{2D}}_{\text{bg}}(\vec r) \left[ n^{\text{2D}}(
\vec {r^\prime} ) - n^{\text{2D}}_{\text{bg}}(\vec {r^\prime} )
\right] \, .
\end{equation}
The term $\delta\mu_{\text{bg}}$, being linear in the charge
distribution related to the excitation, contributes only to the
chemical potential and does not affect the generalized TL model
Hamiltonian because the latter is derived from terms in $\delta
E$ that are quadratic in $\delta n^{\text{2D}}$ and $\delta
n^{\text{2D}}_{\text{bg}}$.

\section{Calculation of sharp-edge characteristic parameters}
\label{append1}

We start with the Hamiltonian of 2D interacting electrons in the
lowest Landau level. After performing the transformation of
particle-hole conjugation, we work consistently in the Fock
space of {\em holes\/} with single-hole states available for
$k\le k_{\text{F0}}$.  This truncation of the Hilbert space
is permitted as long as states with $k$ equal to or in excess of 
$k_{\text{F0}}$ are always occupied by holes. The validity of
this assumption for states close to the sharp-edge ground state
can be verified at the end of the calculation.

Particle-hole conjugation can be performed easily using the
formalism of second quantization. Starting from any operator
expressed in terms of electron creation and annihilation
operators, it is possible to derive its particle-hole conjugate
by replacing the electron's creation operator $c^\dagger_k$
(annihilation operator $c_k$) by the hole's annihilation operator
$h_k$ (creation operator $h^\dagger_k$). Consider the Hamiltonian
for interacting electrons in the lowest Landau level with an
external confining potential present:
\begin{mathletters}\label{electronham}
\begin{eqnarray}
H  &=& H^{0} + H^{\text{int}} \quad ,\\
H^{0} &=& \sum_{k} \, \varepsilon_k \, c_{k}^\dagger c_k
\quad , \\ \label{twobody}
H^{\text{int}} &=& \frac{1}{2L} \sum_{k, p, q} V_q(k-p) \,\,
c_{k+q}^\dagger c_{p}^\dagger c_{p+q} c_k \quad .
\end{eqnarray}
\end{mathletters}
The single-electron dispersion $\varepsilon_k$ is due entirely
to the external potential confining the electrons in the QH
sample, because all electrons in the lowest Landau level have the
same kinetic energy irrespective of their quantum number $k$.
We choose the confining potential to be due to a uniform
background charge that would exactly neutralize the electron
charge if each lowest-Landau-level orbital were occupied with
probability $\nu = 1 - 1/m$:
\begin{equation}\label{bgneutral}
\varepsilon_k = -\nu \sum_{p \le k_{\text{F0}}} \,
V_0 (k - p) \quad .
\end{equation}
Here, $V_q(k-p)$ is the two-body matrix element of the Coulomb
interaction in the Landau-gauge representation of single-particle
states in the lowest Landau level. Explicit expressions for
$V_q(k-p)$ can be found, e.g., in Refs.~\onlinecite{wen:prb:94}
and~\onlinecite{uz:prb:97}. Replacing the electron operators by
hole operators and normal ordering\cite{ahmintro} yields 
\begin{mathletters}\label{holeham}
\begin{eqnarray}
H^* &=& E_{\text{h}} + H^{0}_{\text{h}} +
H^{\text{int}}_{\text{h}} \quad , \\
E_{\text{h}} &=& \sum_{k} \, \Big( \varepsilon_k + \frac{1}{2}
\, \xi_k \Big) \quad , \\
H^{0}_{\text{h}} &=& - \sum_{k} \, \big( \varepsilon_k + \xi_k
\big) \, h_{k}^\dagger h_k \quad , \\
H^{\text{int}}_{\text{h}} &=& \frac{1}{2L} \sum_{k,p,q}
V_q(k-p) \,\, h^\dagger_k h^\dagger_{p+q} h_p h_{k+q} \,\,\, .
\end{eqnarray}
\end{mathletters}
The constant term ($E_{\text{h}}$) in this hole Hamiltonian is
unimportant, but the correction to the single-particle energy
$\xi_k$ plays an essential role in the edge physics:
\begin{equation}\label{implicit}
\xi_k := \frac{1}{L} \, \sum_{p} \, \big[ V_0(k-p) -
V_{k-p}(0) \big] \quad .
\end{equation}

We now evaluate the energy of states where the holes form an
incompressible bulk state with filling factor $1-\nu$ in the
strip for which $y^{\text{(i)}} \le y < \infty$ and form a
filled-Landau-level state in the strip for which $0\le y \le
y^{\text{(o)}}$. The inner strip contributes non-zero occupation
numbers for $k\le k_{\text{F}}^{\text{(i)}}$. Except close to the
edge,\cite{mitra,rm:prb:96} these states are occupied with
probability $1-\nu$. The outer strip contributes non-fluctuating 
integer occupation numbers for states with
$k_{\text{F}}^{\text{(o)}} \le k \le k_{\text{F0}}$. Note that we
have adopted a notation where $k_{\text{F}}^{\text{(o)}}$ is the
inner edge of the outer hole strip. Low-energy excitations can
occur at this edge. In contrast, $ k_{\text{F0}}$ is the outer
edge of the outer hole strip. This edge is formed by the
Hilbert-space truncation and does not support physical
excitations. Our calculations will demonstrate that states of
this type are locally stable. We cannot envisage alternatives and
believe that these states, and their edge-wave excitations, are
the only states in the low-energy portion of the Hilbert space
for sharp edges. 

Since the states we consider have fixed numbers of particles in 
inner and outer strips, it is useful to separate the hole
Hamiltonian into parts as follows: 
\begin{mathletters}
\begin{equation}\label{splitham}
H^* = E_{\text{h}} + H^{\text{(i)}} + H^{\text{(o)}} + \delta H
\quad .
\end{equation}
The term $H^{\text{(i)}}$ describes the inner strip of
interacting holes that is assumed to be confined by a uniform
background neutralizing for {\em holes\/} [density: $(1-\nu)/
(2\pi\ell^2)]$ extending over the interval $y^{\text{(i)}}\le y <
\infty$:
\begin{eqnarray}
H^{\text{(i)}} &=& \sum_{k \le k_{\text{F}}^{\text{(i)}} }
\varepsilon_k^{\text{(i)}} \, h_{k}^\dagger h_k \nonumber \\ &&
+ \frac{1}{2L} \sum_{k, p \le k_{\text{F}}^{\text{(i)}} \atop q}
V_q(k-p) \,\, h^\dagger_k h^\dagger_{p+q} h_p h_{k+q}
\end{eqnarray}
with
$$ \varepsilon_k^{\text{(i)}} := -(1-\nu) \, \frac{1}{L} \,
\sum_{p \le k_{\text{F}}^{\text{(i)}}} V_0(k - p) \quad .$$
This strip is presumed to be in the fractional-QH state at
filling $(1-\nu)$. As it is infinite, the energy per particle
in the inner strip assumes its thermodynamic value\cite{perp}
$\zeta(1-\nu)\approx - 0.41\, e^2/(\epsilon\ell)$. The
contribution $H^{\text{(o)}}$ is for the outer strip of holes,
for which a neutralizing background with density $1/(2\pi
\ell^2)$ is assumed to extend in the region $0\le y \le
y^{\text{(o)}}$. That strip is in the QH state with filling
factor equal to one.
\begin{eqnarray}
H^{\text{(o)}} &=& \sum_{k \ge k_{\text{F}}^{\text{(o)}} }
\varepsilon_k^{\text{(o)}} \, h_{k}^\dagger h_k \nonumber \\
&& + \frac{1}{2L} \sum_{k, p \ge k_{\text{F}}^{\text{(o)}} \atop
q} V_q(k-p) \,\, h^\dagger_k h^\dagger_{p+q} h_p h_{k+q}
\end{eqnarray}
and
$$ \varepsilon_k^{\text{(o)}} := - \frac{1}{L} \, \sum_{p \ge
k_{\text{F}}^{\text{(o)}}} V_0(k - p) \quad .$$
\end{mathletters}
The states we consider have no fluctuations in the quantum
numbers on which $H^{\text{(o)}}$ operates. Since the Hartree
interaction is cancelled by the background, the contribution of
$H^{\text{(o)}}$ to the energy is simply the exchange energy of
the occupied orbitals in the outer strip.

\begin{figure}[hbt]
\epsfxsize3.5in
\centerline{\epsffile{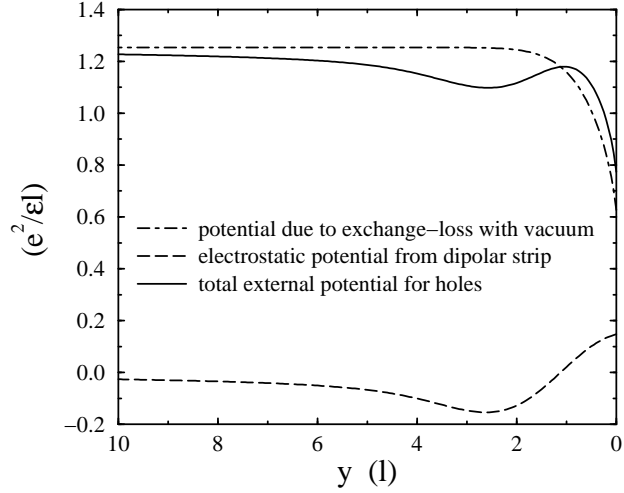}}
\caption{External potential for holes. In addition to the 
electrostatic contribution [dashed curve; cf.\
Eq.~(\ref{hartreedisp})] resulting from the dipolar strip of
charge, there is an additional contribution to the external
potential [dot-dashed curve; cf.\ Eq.~(\ref{fockdisp})] which is
entirely due to particle-hole conjugation in a finite system.
This second contribution attracts holes to the physical edge of
the sample and is essential for the phase separation into an
inner and outer hole strip. The electrostatic potential was
calculated for filling factor $\nu = 2/3$ and a separation
$y^{\text{(i)}} - y^{\text{(o)}} = d_{\text{2/3}} \approx 1.7\,
\ell$ of the inner and outer edges.}
\label{extpot}
\end{figure}

With the terms $H^{\text{(i)}}$ and $H^{\text{(o)}}$ defined
above, Eq.~(\ref{splitham}) constitutes the definition of
$\delta H$. The latter encompasses one-body terms, including
the part from the external potential due to residual background
charge not accounted for in $H^{\text{(i)}} + H^{\text{(o)}}$,
and two-body terms coming from interactions between holes from
different strips. The $q = 0$ interaction terms can be grouped
with the one-body term. The one-body contribution to $\delta H$ 
also contains the exchange contribution to $\xi_k$. In total,
we have
\begin{mathletters}
\begin{eqnarray}
\delta H &=& \delta H_{\text{1-body}}^{\text{eff}} +
\delta H_{\text{2-body}}^{q\neq 0}\\ \label{newdisp}
\delta H_{\text{1-body}}^{\text{eff}} &=& \sum_k
\delta\varepsilon_k \, h_{k}^\dagger h_k \quad ,
\end{eqnarray}
where 
\begin{eqnarray}
\delta\varepsilon_k &=& \delta\varepsilon_k^{\text{H}} +
\delta\varepsilon_k^{\text{F}} \quad , \\ \label{hartreedisp}
\delta\varepsilon_k^{\text{H}} &:=& \Big\{ \sum_{p \ge
k_{\text{F}}^{\text{(o)}}} \frac{\nu}{L} \,\, -
\sum_{k_{\text{F}}^{\text{(i)}} < p < k_{\text{F}}^{\text{(o)}}}
\frac{1-\nu}{L}\Big\} V_0(k-p) \, , \\ \label{fockdisp}
\delta\varepsilon_k^{\text{F}} &:=& \frac{1}{L} \, \sum_p
V_{k-p}(0) \quad .
\end{eqnarray}
\end{mathletters}
The two terms displayed in Eqs.~(\ref{hartreedisp})
and~(\ref{fockdisp}) represent the electrostatic and exchange
contributions to the external potential felt by the holes. In
Fig.~\ref{extpot}, we show their spatial variation. Note that
$\delta \varepsilon_k^{\text{F}}$ appears because of
particle-hole conjugation; it represents the repulsive exchange
interaction between holes and the vacuum which is weaker at the
edge of the system and attracts holes to the physical boundary
of the QH sample. Apart from this term and the constant
$E_{\text{h}}$, the above Hamiltonian could also describe two
strips of \emph{electrons} in the $\tilde\nu = 1/m$ and
$\tilde\nu = 1$ states, respectively. This term is responsible
for the qualitative distinction between the edge structures for
$\nu=1 - 1/m$ and $\nu = 1/m$ bulk fractional-QH states. The $q
\ne 0$ two-body terms give the energy contribution due to
exchange and correlation between electrons in different strips. 

Close to the edge of a QH system that has a filling factor $1/m$
with $m=3,5,\dots$, oscillations occur in the occupation numbers
of the lowest-Landau-level basis states.\cite{mitra} In our model
of a QH edge at filling factor $\nu = 1-1/m$, such oscillations
occur at the inner edge. The expression for the electrostatic
contribution to the external potential which is given in
Eq.~(\ref{hartreedisp}) does not account for the true
occupation-number distribution function at the inner edge.
However, as we comment below, corrections to
Eq.~(\ref{hartreedisp}) are small, and we neglect them.

Now consider the difference in energy between a final state and
an initial state which differ by the transfer of one hole from
the inner strip to the outer one. (See Fig.~\ref{transfer}). We
find that 
\begin{equation}\label{change}
\delta E_{\text{fin}} - \delta E_{\text{ini}} = - \zeta(1-\nu)
- \frac{1}{L} \, \sum_{k = 0}^{k_{\text{F0}} -
k_{\text{F}}^{\text{(o)}}} V_k(0) + \delta E^{\text{res}} \quad .
\end{equation}
The first term in Eq.~(\ref{change}) is the correlation energy
we have to pay to remove the hole from the inner strip, the
second is the exchange energy we gain by putting the hole at the
edge of the outer strip, while the final term contains both
the one-body and two-body contributions from the residual
interaction $\delta E^{\text{res}}$. The one-body piece is
$\delta \varepsilon_{k_{\text{F}}^{\text{(o)}}} - \delta
\varepsilon_{k_{\text{F}}^{\text{(i)}}}$ which can be
interpreted as the change in the self-consistent
(external+Hartree) potential felt by the hole which is being
transferred. If we neglect correlations between holes from
different strips, the two-body residual term consist only of
$\eta_{d}^{\text{(i)}} - \eta_{d}^{\text{(o)}}$ where we denote
the exchange energy for a hole interacting at a distance $h$ with
the inner/outer strips by the symbols $\eta_{h}^{\text{(i)}}$
and $\eta_{h}^{\text{(o)}}$, respectively. Hence we have
\begin{equation}
\delta E^{\text{res}} = \delta
\varepsilon_{k_{\text{F}}^{\text{(o)}}} - \delta
\varepsilon_{k_{\text{F}}^{\text{(i)}}} + \eta_{d}^{\text{(i)}}
- \eta_{d}^{\text{(o)}} \quad .
\end{equation}
Using the expressions
\begin{mathletters}
\begin{eqnarray}
\eta_{d}^{\text{(i)}} &:=& -\frac{1-\nu}{L} \sum_{p \le
k_{\text{F}}^{\text{(i)}}} V_{k_{\text{F}}^{\text{(o)}} - p}
(0) \quad , \\
\eta_{d}^{\text{(o)}} &:=&  - \frac{1}{L} \sum_{p \ge
k_{\text{F}}^{\text{(o)}}} V_{k_{\text{F}}^{\text{(i)}} - p}(0)
\quad ,
\end{eqnarray}
\end{mathletters}
which can be expected to be good for not-too-small distances
$y^{\text{(i)}}_{\text{ini}} - y^{\text{(o)}}_{\text{ini}}\equiv
\big[ k_{\text{F}}^{\text{(o)}} - k_{\text{F}}^{\text{(i)}}\big]
\ell^2$, we find $\delta E_{\text{fin}} - \delta E_{\text{ini}}
= \Delta^{\text{H}} - \Delta^{\text{F}}$, where
\begin{mathletters}
\begin{eqnarray}
\Delta^{\text{H}} &=& \delta
\varepsilon_{k_{\text{F}}^{\text{(o)}}}^{\text{H}} - \delta
\varepsilon_{k_{\text{F}}^{\text{(i)}}}^{\text{H}} \quad , \\
\Delta^{\text{F}} &=& \zeta(1) - \zeta(1-\nu) + \frac{\nu}{L}
\sum_{k\ge k_{\text{F}}^{\text{(o)}} - k_{\text{F}}^{\text{(i)}}}
V_k(0) \, .
\end{eqnarray}
\end{mathletters}

As noted above, our calculation of $\delta E_{\text{fin}} -
\delta E_{\text{ini}}$ neglects contributions due to the
oscillations occurring in the occupation-number distribution
function\cite{mitra} for holes at the inner edge. Taken into
account properly, these oscillations would affect
$\delta E^{\text{res}}$ in essentially the same way as they
affect the energy per particle of the inner QH strip. In
Ref.~\onlinecite{rm:prb:96}, the energy per particle for a
filling factor equal to $1/3$ was calculated for two different
choices of the neutralizing background: (a)~a constant
background-charge density that neutralizes the electron charge in
the bulk, and (b)~a background that neutralizes the electron
charge locally. The difference between the values of the energy
per particle for the models~(a) and (b) corresponds to the
correction to Eqs.~(\ref{hartreedisp}) and (\ref{change}) when
the true occupation-number distribution function is used. This
difference was found\cite{rm:prb:96} to be smaller than $0.0001
\, e^2/(\epsilon \ell)$. The error we make in our calculation of
$\delta E^{\text{res}}$ is therefore three orders of magnitude
smaller than the remaining term in Eq.~(\ref{change}).

Expressions for the matrix elements which are derived for
the Landau gauge\cite{wen:prb:94,uz:prb:96} enable us to
calculate the two contributions $\Delta^{\text{H}}$ and
$\Delta^{\text{F}}$, at least numerically. In Fig.~\ref{numsep},
we show the result for filling factor $\nu = 2/3$. The solid
and dashed curves are the results for $\Delta^{\text{H}}$ and
$\Delta^{\text{F}}$, respectively. In particular, we used
\begin{equation}\label{numhartree}
\delta \varepsilon_k^{\text{H}} = -\frac{e^2}{\epsilon \ell
\pi} \, \frac{1}{\sqrt{2\pi}} \int_{-\infty}^\infty d\kappa \,\,
\kappa \ln |\kappa | \,\, F(\kappa, y / \ell , \lambda/\ell)
\end{equation}
with the definitions $y:= [k_{\text{F0}} - k] \ell^2$ ($\equiv$
coordinate perpendicular to the edge, measured from the physical
edge of the sample towards the bulk), $\lambda :=
y^{\text{(i)}}_1 - y^{\text{(o)}}_1$ ($\equiv$ separation of the
inner and outer edges in the initial state), and
\widetext
\top{-2.8cm}
\begin{equation}
F(\kappa, y, \lambda) := (1-\nu )\, \exp{\left\{ - \frac{[\kappa
- y + \lambda/\nu) ]^2}{2} \right\} } - \exp{\left\{ - \frac{[
\kappa - y + (1-\nu)\lambda/\nu ]^2}{2} \right\} } + \nu \,
\exp{\left\{ - \frac{[ \kappa - y ]^2}{2} \right\} } \quad .
\end{equation}
%\bottom{-2.7cm}
%\narrowtext
%\noindent
To make progress analytically, we have derived a systematic
expansion of $\Delta^{\text{H}}$ in the parameter $\big[
y^{\text{(i)}}_1 - y^{\text{(o)}}_1\big]/\ell$. The asymptotic
result in the limit of large separation of the two edges is
\begin{equation}\label{approxdiff}
\Delta^{\text{H}} = - \frac{e^2}{\epsilon\ell}\,\frac{1}{\pi}
\left[ \nu\ln( \nu ) + (1-\nu) \ln (1-\nu )  \right]
\frac{y^{\text{(i)}}_1 - y^{\text{(o)}}_1}{\ell} \quad ,
\end{equation}
which yields the analytical result for $v_{\text{J}}$ as it is
given in Eq.~(\ref{uliconst}).

\narrowtext

\section{Transverse density profile for edge excitations}
\label{transedge}

Although we use 1D models to describe edge excitations, it is
important to realize that the electrons forming the
fractional-QH sample move in 2D and, therefore, have a wave
function that depends on {\em two\/} coordinates. The part of
the wave function depending on the transverse coordinate ($y$)
is Gaussian with a width of the order of the magnetic length
$\ell$. Hence, the transverse density profile (i.e., the
variation of the 2D density perpendicular to the edge) is not
sharp on scales shorter than $\sim\ell$, even if the
occupation-number distribution function (ONDF) for the
lowest-Landau-level basis states were sharp (as it is the case,
e.g., when the filling factor is equal to one). In this section,
we consider the 2D aspect of edge excitations of fractional-QH
systems at the simple filling factors $\tilde \nu = 1/m$ where
$m=1, 3, \dots$. In particular, the profile of the 2D charge
density perpendicular to the edge is calculated for many-body
states with edge excitations present. The results presented in
this section were applied to the inner and outer {\em hole\/}
strips that arise in the model of a sharp edge of a fractional-QH
sample at filling factor $\nu = 1-1/m$, as discussed above in the
bulk of this article.

The sample geometry considered here is the surface of a
semi-infinite cylinder, see Sec.~\ref{sec3}, which is occupied by
electrons such that the filling factor $\tilde\nu$ is equal to
the inverse of an odd integer. This sample therefore supports a
single branch of edge excitations which are, without loss of
generality, assumed to be right-going. The edge is located at
$y=0$, and the largest wave-vector label of lowest-Landau-level
states that are occupied in the ground state is $k_{\text{F}}$.
To avoid confusion, operators are indicated, in this section, by
a circumflex.

In a symmetric notation, and using our conventions for the sample
geometry, the second-quantized operator of the 2D density in the
lowest Landau level is
\widetext
\top{-2.8cm}
\beq\label{2Ddensity}
\hat n^{\text{2D}}(x, y) = \frac{1}{L} \sum_q \exp\{i q x \} \,
\exp\{-(q\ell)^2/4\} \,\sum_k \frac{\exp\{-(y - [k_{\text{F}} -
k]\ell^2)^2/\ell^2\}}{\pi^{1/2}\,\ell}\,\, c^\dagger_{k+q/2}
c_{k-q/2}\quad .
\eeq
\bottom{-2.7cm}
\narrowtext
\noindent
The operator of the 1D edge density is defined as the integral
of Eq.~(\ref{2Ddensity}) over the transverse coordinate ($y$)
from minus infinity across the edge to a reference point $y = Y
> 0$, located in the bulk:
\beq
\hat\rho^{\text{1D}}(x) = \int_{-\infty}^{Y} dy \,\, \hat
n^{\text{2D}}(x, y) \quad .
\eeq
It is easy to see that the Fourier components of the 1D edge
density operator $\hat\rho^{\text{1D}}(x)$ have the form
\begin{mathletters}
\begin{equation}\label{1Ddensity}
\hat\rho^{\text{1D}}_q = \exp\{-(q\ell)^2 / 4\} \sum_k I_k \,
c^{\dagger}_{k+q/2} c_{k-q/2} \quad ,
\label{rhoq1D}
\end{equation}
where
\begin{equation}
I_k = \frac{1}{\pi^{1/2}\,\ell} \int_{\ell^2k_{\text{F}} -
Y}^{\infty} d y\,\, \exp \{ - (y - \ell^2 k)^2/ \ell^2\} \quad .
\label{I_k}
\end{equation}
\end{mathletters}
As we are interested in the long-wave-length limit $q\ll\ell$
only, the Gaussian prefactor in Eq.~(\ref{1Ddensity}) will be
dropped. In the subspace of low-energy excitations, the Fourier
components of $\hat\rho^{\text{1D}}(x)$ obey the familiar
chiral-Luttinger-liquid commutation relations\cite{wen:int:92}
\beq\label{commute}
\big[\hat\rho^{\text{1D}}_{-q^\prime}\, ,\,\hat\rho^{\text{1D}}_q
\big] = \tilde\nu\,\frac{q L}{2\pi}\,\delta_{q, q^\prime} \quad .
\eeq
Due to the incompressibility of the ground state $\ket{\Psi_0}$
of a fractional-QH system at filling factor $\tilde\nu$, the
operators $\hat\rho^{\text{1D}}_{-q}$ satisfy 
\beq
\hat\rho^{\text{1D}}_{-q}\, \ket{\Psi_0} \equiv 0 \qquad
\mbox{for $q > 0$} \quad .
\eeq

We pose the following problem: Given a state $\ket{\psi}$ in the
edge-excitation subspace that has a 1D density fluctuation
$\delta\varrho(x)$ along the edge, what is the full 2D density
profile for this state? At first sight, this seems like a
question impossible to answer: How can we deduce the 2D density
from its integral over the transverse coordinate? Enabling us to
solve the above problem is the fact that the low-lying
excitations in the system are created by the operators $\hat
\rho^{\text{1D}}_q$ for positive $q$. The edge-density
fluctuation $\delta\varrho(x)$ determines $\ket{\psi}$ uniquely
to be a coherent state\cite{stone:prb:90} of the form
\beq
\ket{\psi} = \exp\left\{ \frac{2\pi}{\tilde\nu L}\sum_{p\ne 0}
\frac{\delta \varrho_{-p}}{p}\,\,\hat\rho^{\text{1D}}_p \right\}
\,\,\,\ket{\Psi_0} \quad .
\eeq
Here, $\delta \varrho_p$ is a Fourier component of the 1D density
fluctuation:
\beq
\delta \varrho_p = \int_0^L d x \,\, e^{i p x}\,\, \delta
\varrho(x) \quad .
\eeq
It is then straightforward to calculate the 2D density
fluctuation $\delta n^{\text{2D}}(x, y)$ associated with the
state $\ket{\psi}$, which is defined by
\beq
\delta n^{\text{2D}}(x, y) = \bra{\psi} \hat n^{\text{2D}}(x, y)
\ket{\psi} - n^{\text{2D}}_0(x, y) \quad ,
\eeq
where we denote the 2D density profile in the ground state by
$n^{\text{2D}}_0(x, y) := \bra{\Psi_0} \hat n^{\text{2D}}(x, y)
\ket{\Psi_0}$. The result is
\beq\label{displace}
\delta n^{\text{2D}}(x, y) = n^{\text{2D}}_0(x, y + 2 \pi
\ell^2\,\delta\varrho(x)/\tilde\nu ) - n^{\text{2D}}_0(x, y)
\quad ,
\eeq
which implies that the 2D density profile for a state with an
edge wave $\delta\varrho(x)$ present differs from the
ground-state density profile by a {\em rigid transverse
deformation\/}. The amount of the transverse displacement is 
$2 \pi\ell^2\,\delta\varrho(x)/\tilde\nu$. Application of this
result to the inner and outer edges of a QH sample at filling
factor $\nu = 1-1/m$ immediately yields Eq.~(\ref{substitute}).

\begin{figure}
\centerline{\epsfxsize3.3in\epsffile{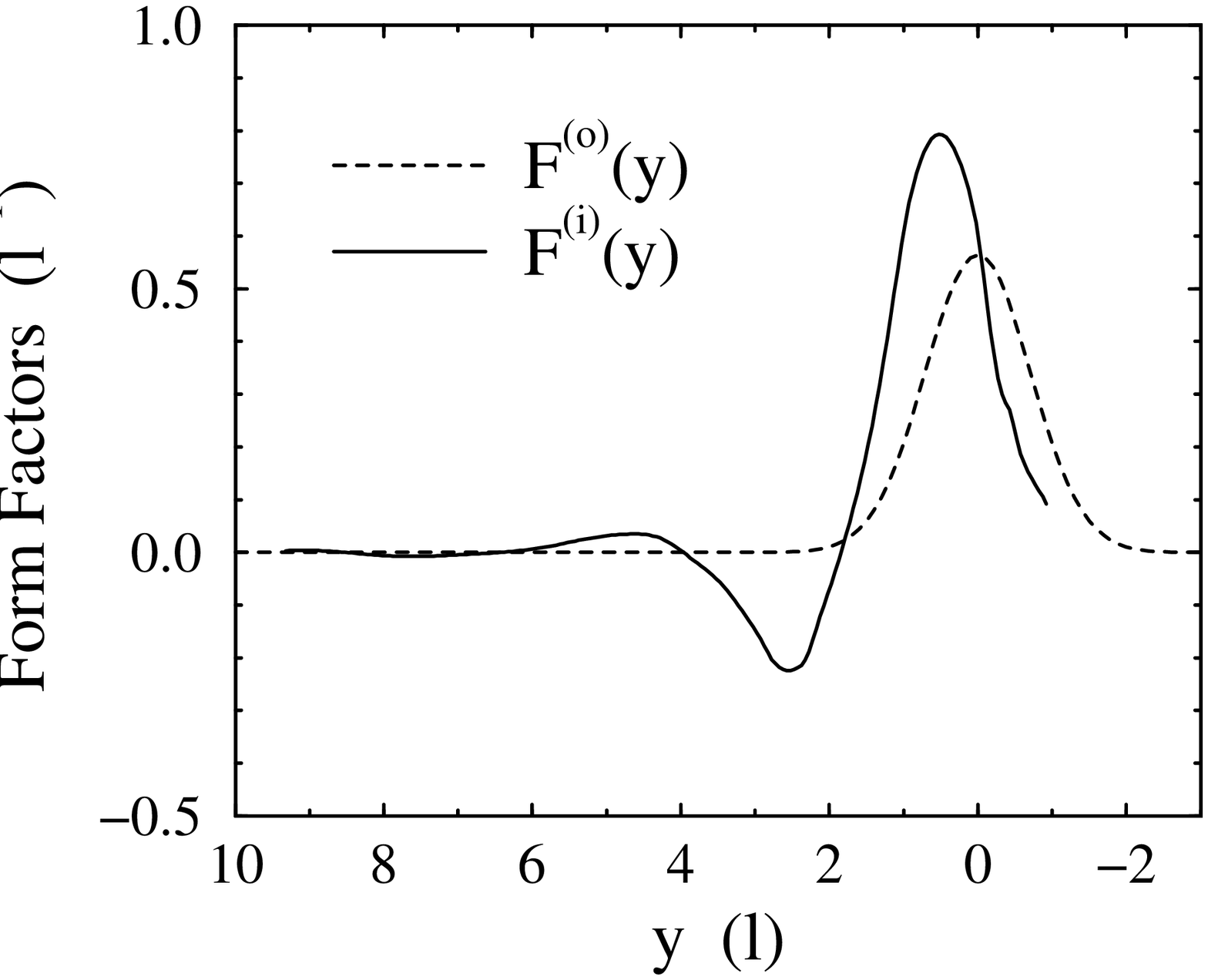}}
\vspace{0.2cm}
\caption{Accounting for the full 2D density profile in our
calculation of the Coulomb contribution, $\delta E_{\text{C}}$,
to the edge-mode energy requires the introduction of appropriate
form factors, $F^{\text{(i)}}(y)$ and $F^{\text{(o)}}(y)$, for
the inner and outer hole strips. (See Sec.~\ref{coulcalc}.) As
shown in Appendix~\ref{transedge}, these form factors are related
to the derivative of the 2D ground-state density profile in the
transverse direction. It is possible to determine $F^{\text{(o)}}
(y)$ analytically [dotted curve, see Eq.~(\ref{outform})] because
there are no fluctuations in the occupation numbers for holes in
the outer strip which has a filling factor equal to one. The
situation is more complicated for the inner hole strip which has
a fractional filling factor equal to $1/m$ with $m=3, 5,\dots$.
As the solid curve, we show the form factor $F^{\text{(i)}}(y)$
for $m=3$ obtained from the ground-state density profile that has
been determined numerically in
Ref.~\protect\onlinecite{rm:prb:96}.}
\label{transprof}
\end{figure}

To determine the parameters in the generalized TL Hamiltonian
describing edge excitations for a QH system at filling factor
$\nu = 1 - 1/m$, we have to calculate the energy $\delta
E_{\text{C}}$ of Eq.~(\ref{coulsil}) up to second order
in the 1D edge-density fluctuations. For that purpose, we need
the 2D density profile of Eq.~(\ref{displace}) only up to first
order in $\delta\varrho(x)$, which reads then
\beq\label{formfact}
\delta n^{\text{2D}}(x, y) = \frac{2\pi\ell^2}{\tilde\nu}\,\,
\left[ \partial_y n^{\text{2D}}_0(x, y) \right] \,\, \delta
\varrho(x) \quad .
\eeq
In a situation where the ONDF is a step function with a step of
height $\tilde\nu$ at $k = k_{\text{F}}$, one finds the
analytical result
\beq\label{oneresult}
\frac{2\pi\ell^2}{\tilde\nu} \, \partial_y n^{\text{2D}}_0(x, y)
= \frac{\exp\left( - y^2 /\ell^2 \right)}{\sqrt{\pi}\ell} \quad .
\eeq
Equation~(\ref{oneresult}) is exact for a QH strip at a filling
factor equal to one. It also applies to the density profile of
the neutralizing background we have chosen
[see Eq.(\ref{bgneutral})]. We can then deduce the form factors
to be used in Eqs.~(\ref{2Dcharge}); they are
\begin{mathletters}\label{formfactors}
\begin{eqnarray}
F^{\text{(i)}}(y) &=& \frac{2\pi\ell^2}{1 - \nu}\,\, \partial_y
n^{\text{(i)}}_0\big(x, y - y^{\text{(i)}}\big) \quad , \\
\label{outform}
F^{\text{(o)}}(y) &=& \frac{\exp\left( - y^2 /\ell^2 \right)}
{\sqrt{\pi}\ell} \quad , \\
F^{\text{(b)}}(y) &=& \frac{\exp\left( - y^2 /\ell^2 \right)}
{\sqrt{\pi}\ell} \quad .
\end{eqnarray}
\end{mathletters}
We have denoted the 2D ground-state {\em hole\/} density for the
inner strip by $n^{\text{(i)}}_0$. At present, it is not possible
to give a closed-form analytical result for $F^{\text{(i)}}(y)$.
So far, the 2D density profile and ONDF for fractional-QH systems
with $\tilde\nu = 1/3$, $1/5$, and $1/7$ have only been obtained
numerically for small numbers of particles.\cite{mitra,rm:prb:96}
It is established that the ONDF in fractional-QH systems at the
simple $1/m$ filling factors is not a step
function.\cite{wen:int:92} With a broadened ONDF at the inner
edge, we also expect $F^{\text{(i)}}(y)$ to be broader than
$F^{\text{(o)}}(y)$. However, the form factor $F^{\text{(i)}}(y)$
differs from $F^{\text{(o)}}(y)\equiv F^{\text{(b)}}(y)$ in a
more significant way because oscillations
appear\cite{mitra,rm:prb:96} in the ONDF and the 2D density
profile close to the edge of a $\tilde\nu=1/m$ QH sample when
$m>1$. In the long-wave-length limit, all these effects are taken
account of in the correction factors $\Upsilon_\lambda$ and
$\Upsilon_{\text{C}}$. [See Eqs.~(\ref{somecoul}).] To compute
actual numbers for the experimentally most relevant case of
$\nu = 2/3$, we have taken the data reported in Fig.~3 of
Ref.~\onlinecite{rm:prb:96} for the 2D ground-state density
profile of a fractional-QH system at $\tilde\nu = 1/3$ and
derived the corresponding form factor $F^{\text{(i)}}(y)$. The
result is given in Fig.~\ref{transprof}, where we also show
$F^{\text{(o)}}(y)$ as it is determined from Eq.~(\ref{outform}).

Using the analytical expressions for $F^{\text{(o)}}(y) = 
F^{\text{(b)}}(y)$ and the numerical result for
$F^{\text{(i)}}(y)$ as shown in Fig.~\ref{transprof}, we
determine the the wave-vector-independent quantities
$\Delta^{\text{(bi)}}$, $\Delta^{\text{(bo)}}$, and
$\Delta^{\text{(bb)}}$ for the case of $d = d_{2/3} \approx
1.7~\ell$. This yields Eqs.~(\ref{somecoul}) with the quoted
values of the correction factors.

\widetext

\end{appendix}

%\pagebreak

\begin{thebibliography}{10}
\bibitem{realspace}
Here and in the following, $y^{\text{(i)}}$ is the coordinate
which the outermost lowest-Landau-level orbital from the inner
strip is centered at. Likewise, the orbital with the smallest
wave-vector label among the states occupied by holes from the
outer hole strip is centered at $y^{\text{(o)}}$.
\bibitem{epsilon}
The appropriate value of $\epsilon$ depends on the specific
sample geometry in which the 2D electron system is realized and
need not be the bulk-semiconductor dielectric constant
$\epsilon_{\text{GaAs}}\approx 13$. Only if the wave length of
the excitations of interest is much smaller than the distance of
the 2D electron system to the surface of the (3D) semiconductor
sample, $\epsilon = \epsilon_{\text{GaAs}}$. The opposite limit
is, however, more typical. We therefore use $\epsilon = (1 +
\epsilon_{\text{GaAs}})/2$ for the numerical estimates presented
in this work.
\bibitem{perp}
The energy per particle $\zeta(\gamma)$ for fractional-QH
systems at filling factor $\gamma$ has been calculated for
various fillings by exact numerical
diagonalization,\cite{rm:prb:96,dai:prb:84,kas:ann:94}
Monte-Carlo methods,\cite{rm:prb:86,rm:prb:96} and using the
variational wave function proposed by
Laughlin.\cite{rbl:prl:83} To determine the parameter $d$,
we need to know $\zeta(1)$ and $\zeta(1-\nu)$. Simply from a
calculation of the exchange energy for an electron at the edge
of a filling-factor-one QH strip, we know\cite{ahm:prb:86}
that $\zeta(1) =-\sqrt{\pi/8} \,\, e^2 / (\epsilon \ell)$.
In our numerical calculations for the case of $\nu=2/3$, we used
$\zeta(1/3) = - 0.41\, e^2/ (\epsilon \ell)$.
\end{thebibliography}
\end{document}